\documentclass[useAMS,usenatbib]{mn2e}
\usepackage{times}
\usepackage{amssymb}
\usepackage{graphicx}
\usepackage{rotating}
\usepackage{lscape}
\usepackage{multirow}

\begin{document}

% If your system does not have the AMS fonts version 2.0 installed, then
% remove the useAMS option.
%
% useAMS allows you to obtain upright Greek characters.
% e.g. \umu, \upi etc.  See the section on "Upright Greek characters" in
% this guide for further information.
%
% If you are using AMS 2.0 fonts, bold math letters/symbols are available
% at a larger range of sizes for NFSS release 1 and 2 (using \boldmath or
% preferably \bmath).
%
% The usenatbib command allows the use of Patrick Daly's natbib.sty for
% cross-referencing.
%
% If you wish to typeset the paper in Times font (if you do not have the
% PostScript Type 1 Computer Modern fonts you will need to do this to get
% smoother fonts in a PDF file) then uncomment the next line
% \usepackage{Times}

%%%%% AUTHORS - PLACE YOUR OWN MACROS HERE %%%%%

%%%%%%%%%%%%%%%%%%%%%%%%%%%%%%%%%%%%%%%%%%%%%%%%

\title[The sizes of Brightest Cluster Galaxies]{Little change in the sizes of the most massive galaxies since $z=1$} 
\author[J.P. Stott et al.]{J.P. Stott$^{1, 2}$\thanks{E-mail:
j.p.stott@durham.ac.uk},  C. A. Collins$^{2}$, C. Burke$^{2}$, V. Hamilton-Morris$^{3}$ and G. P. Smith$^{3}$
%\footnotemark[1]\thanks{This file has been amended to
%highlight the proper use of \LaTeXe\ code with the class file.
%These changes are for illustrative purposes and do not reflect the
%original paper by A. V. Raveendran.}
\\
$^{1}$Extragalactic \& Cosmology Group, Department of Physics, University of Durham, South Road, Durham DH1 3LE, UK\\
$^{2}$Astrophysics Research Institute, Liverpool John Moores University, 12 quays House, Egerton Wharf, Birkenhead, UK\\
$^{3}$School of Physics \& Astronomy, University of Birmingham, Edgbaston, Birmingham, B15 2TT, UK\\}

%$^{2}$Building, Institute, Street Address, City, Code, Country}

\date{}

\pagerange{\pageref{firstpage}--\pageref{lastpage}} \pubyear{2002}

\maketitle

\label{firstpage}

\begin{abstract}

Recent reports suggest that elliptical galaxies have increased their size dramatically over the last $\sim8$ Gyr. This result  points to a major re-think of the processes dominating the late-time evolution of galaxies. In this paper we present the first estimates for the scale sizes of brightest cluster galaxies (BCGs) in the redshift range $0.8 < z < 1.3$ from an analysis of deep {\it Hubble Space Telescope} imaging, comparing to a well matched local sample taken from the Local Cluster Substructure Survey at $z\sim0.2$. For a small sample of 5 high redshift BCGs we measure half-light radii ranging from $14-53$\,kpc using de Vaucuoleurs profile fits, with an average determined from stacking of $32.1\pm2.5$\,kpc compared to a value $43.2\pm 1.0$\,kpc for the low redshift comparison sample. This implies that the scale sizes of BCGs at $z=1$ are $ \simeq30\%$ smaller than at $z=0.25$. Analyses comparing either S\'{e}rsic  or Petrosian radii also indicate little or no evolution between the two samples. The detection of only modest evolution at most out to $z=1$ argues against BCGs having undergone the large increase in size reported for massive galaxies since $z=2$ and in fact the scale-size evolution of BCGs appears closer to that reported for radio galaxies over a similar epoch. We conclude that this lack of size evolution, particularly when coupled with recent results on the lack of BCG stellar mass evolution, demonstrates that major merging is not an important process in the late time evolution of these systems. The homogeneity and maturity of BCGs at $z=1$ continues to challenge galaxy evolution models.
 
\end{abstract}

\begin{keywords}
galaxies: clusters: general -- galaxies: evolution -- galaxies: elliptical and lenticular, cD
\end{keywords}

\section{Introduction}

There is currently great interest in the evolution of the scale size of massive galaxies over the age of the Universe, as this provides constraints on the merger history and feedback processes that govern galaxy evolution. However, this is still a controversial field as there have been claims that massive galaxies at high redshift are smaller than galaxies with a similar mass in the local Universe \citep{Daddi2005,Trujillo2006,Trujillo2007,Toft2007,cimatti2008,franx2008,vanderwel2008,vandokkum2008,damjanov2009,vandokkum2009,vandokkum2010}; however others reach different conclusions (e.g. \citealt{mancini2010}) and one major criticism is that any systematic errors involved in the high redshift observations and analysis would tend to bias a result towards such an evolution \citep{driver2010}. In addition, there are observations of massive and compact galaxies in the local Universe and therefore, rather than constituting an evolution, the massive compact galaxies seen at high redshift may have remained unchanged and still exist today \citep{Valentinuzzi2010, saracco2010}.

In this paper we look for evidence of a size evolution in brightest cluster galaxies (BCGs), the most luminous and homogenous class of galaxy \citep{sandage1972,sandage1976,cm98,whey08,stott08,collins09}. The advantage of studying BCGs in particular, is that they are also the largest galaxies and can therefore be seen at great distances; but most importantly, they reside at the centres of rich clusters which are thought to be comparable environments across all epochs.

There have been a large number of studies of the surface brightness profiles and scale sizes of BCGs in the local Universe (e.g. \citealt{oemler1976, carter1977,hoessel1987,schombert1986, schombert1987,oegerle1991,graham1996,bernardi2007,vonderlinden2007,fasano2009}) with measured scale sizes found to range typically from $\sim$10kpc to $\sim$100kpc.
In terms of evolution in the scale size of BCGs, \cite{bernardi2009} find that when comparing samples in an internally consistent way there is a 70\% increase in the size of BCGs in the 3\,Gyr between $z=0.25$ and $z\sim0$, thought to be due to dry, dissipationless minor merging. \cite{ascaso2010} also find an increase in size between $z\sim0.4$ and $z\sim0$, although with no measured change in the shape of the profile, which they claim cannot therefore be due to merging and is thus caused by feedback processes in the centre of the galaxy.

Measuring the scale size of large galaxies and BCGs in particular is notoriously difficult (see \citealt{Lauer2007}) and one of the principle concerns of interpreting the scale-size evolution of BCGs and elliptical galaxies is the significant disparity in published values even at low redshift. For example, the scale radii measured by \cite{vonderlinden2007} are significantly smaller than those measured by both \cite{bernardi2007} and \cite{liu2008} for essentially the same Sloan Digital Sky Survey (SDSS) data. At moderate redshift there is also significant disagreement. As an illustration one particular BCG (in the cluster MS\,0451$-$03) with a scale size measured using a de Vaucouleurs profile from near-IR HST data by \cite{nelson2002b} is found to be a factor 6 smaller than a similar optical analysis of the same galaxy by \cite{bildfell2008}; in this case possibly due to the small field of view of the \cite{nelson2002b} observation and the presence of a foreground spiral galaxy dominating the light at larger BCG-centric radii. In other cases these differences appear to be due to the subtly different ways that the scale size of BCGs is defined or measured and also the variable depth and quality of the data between samples. Some authors fit only de Vaucoleur profiles while others fit free S\'{e}rsic curves, which can give significantly different results, and in some cases double fits (e.g. S\'{e}rsic $+$ exponential, \citealt{ascaso2010}) are used which can be difficult to interpret or compare with any other investigations. In addition, some studies only fit circularly symmetric profiles (e.g. \citealt{bildfell2008}) while others include ellipticity (e.g. \citealt{liu2008}) which can give different results depending on the galaxy's shape \citep{nelson2002b}. There is also the problem that S\'{e}rsic index and scale size are coupled (e.g \citealt{graham1996}) and as such when fitting a free S\'{e}rsic curve one can find answers that are degenerate in these two parameters. In terms of the data, although the S\'{e}rsic fit is parametric and therefore depends only on the shape of the surface brightness profile, the depth of the data or the limiting surface brightness used for the fitting process may in some studies be inadequate to see enough of the profile to return a good description of the galaxy's light distribution, shown to be an important factor by \cite{liu2008} when explaining the discrepancy between their results and those of \cite{vonderlinden2007}. Different treatments of the background subtraction, particularly at longer wavelengths, and the masking out of near neighbours in crowded fields may also lead to a discrepancy between results from different studies.  In addition, significant atmospheric seeing from ground-based measurements, particularly when going to moderate redshifts, if not correctly deconvolved with the galaxy image, may cause erroneous fits to the inner parts of galaxies affecting the entire measured profile.

BCGs are generally found to lie off the Kormendy relation (surface brightness versus scale size, \citealt{kormendy1977}), having larger radii for a given surface brightness compared to the general elliptical galaxy population \citep{hoessel1987,schombert1987, oegerle1991}. A similar relation to this, luminosity versus scale size, is found to be steeper for BCGs than for regular ellipticals \citep{bernardi2007,liu2008}, although other authors using essentially the same data have not seen evidence for a steeper slope \citep{vonderlinden2007}. 
The offset of BCGs from the Kormendy relation of regular ellipticals is most likely due to the extended cD halo associated with BCGs in rich clusters, which is a low surface brightness component of the galaxy light profile \citep{richstone1976,lopezcruz1997}. The origin of this halo is thought to be the addition of tidally stripped material, from other cluster members, to the outer regions of the BCG \citep{hoessel1987}. The halo itself however, provides difficulties for the study of BCG sizes as this is sometimes found to be a separate component of the surface brightness, potentially requiring a double fit, or it can modify a single fit to be more like a power law (S\'{e}rsic index $n\rightarrow\infty$, \citealt{schombert1986}) rather than a de Vaucouleur profile (S\'{e}rsic index, $n=4$), which therefore does not converge and can continue into the intra-cluster light (ICL) for 100s kpc \citep{carter1977, oemler1976}. Estimates suggest as much as $33-40\%$ of an entire cluster's optical luminosity is contained in the BCG $+$ ICL of which $80\%$ is in the ICL \citep{gonzalez2007}. To account for this many investigators fit BCG light profiles to a consistent surface brightness level (e.g. \citealt{graham1996}) as this has been shown to significantly affect the recovered value for the galaxy size \citep{liu2008}

Our search for evolution in the scale-size of BCGs is timely considering our recent work which suggests that BCGs as a population undergo little evolution in their stellar mass out to $z=1.5$ \citep{collins09,stott2010}. Any evolution we see in the scale size would suggest a number of possible processes dominating the late-time evolution of BCGs: minor merging; interactions which add mass to the outer regions of the galaxy; feedback from an active galactic nuclei (AGN); or a central starburst that disrupts the mass in the centre of the systems giving rise to adiabatic expansion \citep{hopkins2010}. In any case from a theoretical point of view,  brightest cluster galaxies should be ideal candidates with which to study size and shape evolution as a result of  hierarchical assembly, because they are thought to have undergone many mergers and interactions over their history as a result of being located so close to the centre of mass of the largest dark matter haloes \citep{deluc07}. 

This paper sets out the first comprehensive study of BCG scale sizes in a small  X-ray selected sample at high redshift ($z>0.8$) and compares the results to a similar and larger sample in the local Universe ($z\sim0.2$), exploiting the best space-based imaging data available, well matched in both rest-frame filter and depth, in an attempt to reduce the problems which beset other studies discussed above.

A Lambda Cold Dark Matter ($\Lambda$CDM) cosmology ($\Omega_{M}=$0.3, $\Omega_{Vac}=$0.7, $H_{0}=$70 km s$^{-1}$ Mpc$^{-1}$) is used throughout this work.

\section[]{The Sample}

Table \ref{tab:sample} details our sample of 5 of the most distant, spectroscopically confirmed, galaxy clusters. The sample consists of clusters discovered by several X-ray surveys. The clusters all have redshifts in the range $0.8 < z < 1.3$ and have X-ray luminosities of $ 1\lesssim L_{X}\lesssim18\times10^{44}\rm erg\, s^{-1}$ ($0.1 - 2.4$ keV). All of these clusters have been imaged with the {\it Hubble Space Telescope} Advanced Camera for Surveys (HST/ACS) through the F850LP filter with the exception of CL\,J1226\,$+$\,3332 which is imaged with the F814W filter. We choose these particular 5 clusters as they are the highest redshift clusters with the longest image exposure times ($>10\rm ks$) which is key for observations of low surface brightness features  that may otherwise be difficult to detect due to the strong dependancy of cosmological surface brightness dimming on redshift. The data for CL\,J0152.7\,$-$\,1357, MS1054.4\,$-$\,0321, RDCS J0910 $+$ 5422 and RDCS\,J1252.9\,$-$\,2927 are from Cycle 11 proposal ID 9290 and CL\,J1226\,$+$\,3332 is from Cycle 10 proposal ID 9033 and are downloaded from HST archive in reduced form. The reason we choose to analyse HST data rather than ground based observations is because the affects of atmospheric seeing are removed with only the relatively small instrumental point spread function (PSF) to consider, which makes fitting the profile shape considerably easier at small radii and with no near-IR sky background to contend with these HST observations are thus ideal for this project. 

%We define a low redshift comparison sample of 19 clusters (with median $z=0.23$) sourced from the Local Cluster Substructure Survey (LoCuSS, \citealt{smith2010}), detailed in Table \ref{tab:sampleloz}. These clusters were all imaged in the HST ACS in F606W filter which gives good rest frame filter agreement with our high redshift F850LP sample. The majority of these data are from HST Cycle 15, proposal ID 10881 (P.I.: G. P. Smith) with an additional 3 archival clusters from Cycle 11 proposal ID 9270 and 2 from Cycle 13 proposal ID 10200 (the 2 components of the Bullet Cluster). These data were reduced using the standard STScI MultiDrizzle software (further details of the sample and data reduction to appear in Hamilton-Morris et al. in prep.). Each of the clusters was observed for 1200s. We choose to analyse our own low redshift sample rather than compare to those in the literature due to the confusion about methods of effective radius measurement and the various resulting discrepancies seen between studies at low $z$ discussed in the introduction. The average X-ray luminosity of the low redshift clusters in the $0.1-2.4 \rm \,keV$ band is $\sim7\times10^{44}\rm erg s^{-1}$ with an average X-ray temperature of 7\rm\,keV which is similar to that of their high redshift counterparts (average $\rm L_{X}\sim6\times10^{44}\rm erg s^{-1}$ and $\rm T_{X}\sim7\,keV$) and as these correlate with cluster mass, albeit with a significant scatter (e.g. \citealt{randb2002,vik09,hoekstra2011}), these can be thought of as well matched samples.

We define a low redshift comparison sample of 19 clusters at $z_{\rm median}=0.23$ for which archival \emph{HST}/ACS data are available (Table \ref{tab:sampleloz}).  These are a subset of the clusters selected for study by the Local Cluster Substructure Survey (LoCuSS, \citealt{smith2010}) from the ROSAT All-sky Survey catalogues \citep{ebeling98,ebeling2000,bohringer2004}.  These clusters were all observed with HST/ACS through the F606W filter which gives good rest frame filter agreement with our high redshift F850LP sample. The majority of these data were acquired in Cycle 15 (PID:10881; P.I.: G.\ P.\ Smith) with the balance from Cycle 11 (A\,611, Z2701, A\,2537; PID:9270; P.I.: S.\ W.\ Allen) and  Cycle 13 (PID:10200; 1ES0657-558; P.I.: C.\ Jones).  All of the observations employed standard dither patterns, and spanned at least 1200sec.  The flat-fielded frames were reduced using standard STScI MultiDrizzle routines (Hamilton-Morris et al., in prep.). We choose to analyse our own low redshift sample rather than compare to those in the literature due to the confusion about methods of effective radius measurement and the various resulting discrepancies seen between studies at low-$z$ discussed in the introduction. The average X-ray luminosity of the low redshift clusters in the $0.1-2.4 \rm \,keV$ band is $\sim7\times10^{44}\rm erg s^{-1}$ with an average X-ray temperature of $6.5\rm \pm2.2\,keV$ which is similar to that of their high redshift counterparts (average $\rm L_{X}\sim6\times10^{44}\rm erg s^{-1}$ and $\rm T_{X}=7.2\pm1.8\,keV$) and as these properties correlate with cluster mass, albeit with a significant scatter (e.g. \citealt{randb2002,maughan2006,vik09,hoekstra2011}), these can be thought of as well matched samples.

\begin{table*}
\begin{center}
%\caption[]{The high redshift cluster sample. The radii quoted are along the semi-major axis. $b/a$ is the axis ratio of the BCG. $\chi^{2}_{\nu}$ is the reduced chi-square statistic for each fit.}
\caption[]{The high redshift cluster sample. The radii quoted are along the semi-major axis. $b/a$ is the axis ratio of the BCG. }
\label{tab:sample}
\small\begin{tabular}{llllllllll}

\hline
%Cluster & R.A.\ & Dec.\ & $z$  &$r_{e} (n=4)$&$\chi^{2}_{\nu}$&$r_{e}$ (free)&$n$&$\chi^{2}_{\nu}$&$b/a$\\
%&\multicolumn{2}{c}{(J2000)}&&(kpc) &&(kpc)&\\
Cluster & R.A.\ & Dec.\ & $z$  &$r_{e} (n=4)$&$r_{e}$ (free)&$n$&$b/a$\\
&\multicolumn{2}{c}{(J2000)}&&(kpc) &(kpc)&\\
\hline

%CL J0152.7 $-$ 1357				&01 52 41.0&$-$13 57 45	&0.83		&27.2$\pm$0.4&5.1		&16.8$\pm$0.5&2.9$\pm$0.1&1.6	&0.90\\
%RDCS J0910 $+$ 5422			&09 10 44.9&54 22 09&1.11 				&14.4$\pm$0.6&1.8		&8.5$\pm$0.5&2.5$\pm$0.2&0.7	&0.73\\
%MS1054.4 $-$ 0321				&10 57 00.2&$-$03 37 27	&0.82			&24.2$\pm$0.3&11.3	&62.6$\pm$5.2&6.7$\pm$0.2&1.4	&0.63\\
%CL J1226 $+$ 3332				&12 26 58.0&$+$33 32 54	&0.89		&52.6$\pm$0.5&3.1		&59.4$\pm$2.0&4.3$\pm$0.1&2.8	&0.70\\	
%RDCS J1252.9 $-$ 2927			&12 52 54.4&$-$29 27 17	& 1.24			&29.7$\pm$0.5&6.3		&80.3$\pm$11.9&6.8$\pm$0.4&1.5	&0.80\\
CL J0152.7 $-$ 1357				&01 52 41.0&$-$13 57 45	&0.83		&27.2$\pm$0.4	&16.8$\pm$0.5&2.9$\pm$0.1	&0.90\\
RDCS J0910 $+$ 5422			&09 10 44.9&54 22 09&1.11 				&14.4$\pm$0.6		&8.5$\pm$0.5&2.5$\pm$0.2	&0.73\\
MS1054.4 $-$ 0321				&10 57 00.2&$-$03 37 27	&0.82			&24.2$\pm$0.3	&62.6$\pm$5.2&6.7$\pm$0.2	&0.63\\
CL J1226 $+$ 3332				&12 26 58.0&$+$33 32 54	&0.89		&52.6$\pm$0.5		&59.4$\pm$2.0&4.3$\pm$0.1	&0.70\\	
RDCS J1252.9 $-$ 2927			&12 52 54.4&$-$29 27 17	& 1.24			&29.7$\pm$0.5	&80.3$\pm$11.9&6.8$\pm$0.4	&0.80\\

\hline
\end{tabular}
\end{center}
\end{table*}

\begin{table*}
\begin{center}
%\caption[]{The low redshift comparison sample. The radii quoted are along the semi-major axis. $b/a$ is the axis ratio of the BCG. $\chi^{2}_{\nu}$ is the reduced chi-square statistic for each fit.}
\caption[]{The low redshift comparison sample. The radii quoted are along the semi-major axis. $b/a$ is the axis ratio of the BCG.}
\label{tab:sampleloz}
\small\begin{tabular}{llllllllll}

\hline
%Cluster & R.A.\ & Dec.\ & $z$  &$r_{e} (n=4)$&$\chi^{2}_{\nu}$&$r_{e}$ (free)&$n$&$\chi^{2}_{\nu}$&$b/a$\\
%&\multicolumn{2}{c}{(J2000)}&&(kpc) &&(kpc)&\\
Cluster & R.A.\ & Dec.\ & $z$  &$r_{e} (n=4)$&$r_{e}$ (free)&$n$&$b/a$\\
&\multicolumn{2}{c}{(J2000)}&&(kpc) &(kpc)&\\

\hline

	Abell\,2813  	&00 43 25.5 	&$-$20 37 02 			&0.292	&87.0$\pm$1.3&	52.5$\pm$1.9&3.1$\pm$0.1&		0.59\\
	
	%RXCJ0105.5 $-$ 2439 	&01 05 37.6 	&$-$24 40 49 	& 0.23	&49.0$\pm$0.2&&	24.8$\pm$0.2&2.8\\	
	Abell\,141  	&01 05 37.6 	&$-$24 40 49 			& 0.230	&55.3$\pm$0.4&	34.6$\pm$0.7&3.2$\pm$0.1&		0.65\\
	
	%RXCJ0118.1 $-$ 2658 	&01 18 11.5 	&$-$26 58 11 	&0.227   	&32.6$\pm$0.1&&	49.3$\pm$0.9&4.8\\
	Abell\,2895	&01 18 11.5 	&$-$26 58 11 			&0.227   	&38.8$\pm$0.2&	65.8$\pm$1.8&4.9$\pm$0.1&		0.76\\
	
	RXCJ0220.9 $-$ 3829 	&02 20 56.5 	&$-$38 28 54 	&0.229	&35.8$\pm$0.3&	30.8$\pm$0.6&3.7$\pm$0.1&		0.74\\
	
	%RXCJ0237.4 $-$ 2630 	&02 37 28.2 	&$-$26 30 29    &0.22	&78.4$\pm$0.6&&	145.3$\pm$5.1&5.0\\
	Abell\,368 	&02 37 28.2 	&$-$26 30 29    		&0.22	&77.1$\pm$0.8&	129.4$\pm$8.9&4.9$\pm$0.1&	0.52\\
	
	RXCJ0304.1 $-$ 3656 	&03 04 04.4 	&$-$36 56 28 	& 0.219	&24.6$\pm$0.1&	95.4$\pm$4.3&7.3$\pm$0.1&	0.86\\
	%Abell\,3084 	&03 04 04.4 	&$-$36 56 28 			& 0.219	&24.5$\pm$0.1&&	133.6$\pm$4.8&8.1&0.86\\
	
	RXCJ0336.3 $-$ 4037 	&03 36 16.4 	&$-$40 37 44 	&0.173	&12.1$\pm$0.1&	376.5$\pm$72.8&12.9$\pm$0.5&	0.95\\
	%Abell\,3140 	&03 36 16.4 	&$-$40 37 44 			&0.178	&11.4$\pm$0.1&&	441.3$\pm$55.4&15.5&0.95\\
	
	%RXCJ0358.8 $-$ 2955 	&03 58 53.7 	&$-$29 55 36 	&0.1681	&29.7$\pm$0.4&&	118.8$\pm$13.2&6.6\\
	%Abell\,3192 	&03 58 53.7 	&$-$29 55 36 			&0.168	&30.8$\pm$0.4&11.8&	120.7$\pm$14.5&6.5$\pm$0.2&3.6&	0.46\\
	
	RXCJ0449.9 $-$ 4440 	&04 49 56.9 	&$-$44 40 21 	&0.1501	&57.9$\pm$0.3&	40.8$\pm$0.8&3.3$\pm$0.1&		0.67\\
	%Abell\,3292 	&04 49 56.9 	&$-$44 40 21 			&0.150	&55.8$\pm$0.2&&	33.7$\pm$0.2&3.1&0.67\\
	
	RXCJ0528.2 $-$ 2942 	&05 28 15.4 	&$-$29 42 59 	&0.158	&60.7$\pm$0.4&	25.8$\pm$0.3&2.6$\pm$0.1&		0.59\\
	
	%RXCJ0547.6 $-$ 3152 	&05 47 38.1 	&$-$31 52 20 	&0.1483	&14.1$\pm$0.1&&	33.5$\pm$0.6&6.3\\
	Abell\,3364	&05 47 38.1 	&$-$31 52 20 			&0.148	&15.7$\pm$0.1&	45.4$\pm$1.7&6.4$\pm$0.1&		0.95\\
	
	RXCJ0638.7 $-$ 5358 	&06 38 45.8 	&$-$53 58 23 	&0.227	&30.1$\pm$0.2&	95.9$\pm$5.2&6.9$\pm$0.1&	0.72\\
	
	GC065819	&06 58 15.9&$-$55 56 36				& 0.297	&54.1$\pm$0.5&	254.6$\pm$25.2&6.6$\pm$0.2&	0.71\\
	
	GC065832	&06 58 38.1 &$-$55 57 25			& 0.296	&19.8$\pm$0.2&	$\rightarrow\infty$&41.0$\pm$6.0&	0.73\\
	
	Abell\,611		&08 00 56.8&$+$36 03 24		&	 0.288	&66.5$\pm$0.3&	27.7$\pm$0.3&2.5$\pm$0.1&		0.70\\
	
	Abell\,781 			&09 20 26.8 	&$+$30 30 38 	&0.29	&27.4$\pm$0.2&	24.3$\pm$0.8&3.8$\pm$0.1&		0.88\\
	
	Zwicky\,2701		&09 52 49.1&$+$51 53 06		& 0.214	&43.8$\pm$0.3&	34.4$\pm$0.5&3.6$\pm$0.1&		0.68\\
	
	RXJ1000.5 $+$ 4409 	&10 00 31.3 	&$+$44 08 48 	&0.154	&12.4$\pm$0.1&	123.8$\pm$13.6&10.0$\pm$0.3&		0.81\\

	Abell\,2187  			&16 24 14.0  	&$+$41 14 31  	&0.183	&48.1$\pm$0.3&	38.0$\pm$0.5&3.6$\pm$0.1&		0.68\\	 

	Abell\,2537		&23 08 22.3&$-$02 11 32			& 0.295	&56.7$\pm$0.3&	31.9$\pm$0.5&3.1$\pm$0.1&		0.73\\

\hline
\end{tabular}
\end{center}
\end{table*}

\subsection{BCG selection}
The identification of BCGs within clusters is usually obvious from visual inspection of the images since, for such rich clusters, they are the prominent galaxy closest to the X-ray centroid, often with a cD-like profile. However for our high redshift sample we chose to formalize this by studying the tip of the red sequence in the colour-magnitude relations given in  \cite{stott2010}.  
For each cluster we identified the red sequence with $J-K_{s}$ colour and selected the brightest galaxy from the $K_{s}$-band magnitudes of all the red sequence galaxies within a projected distance of 500~kpc from the cluster X-ray centroid. \cite{lm04} have shown that approximately $95\%$ of massive clusters the BCG lies within this radius. Furthermore these are all spectroscopically confirmed members of their host clusters. The BCG selection is more straightforward at low-redshift, because the ACS observations are centered on the spectroscopically confirmed BCG, many of which lie adjacent to strongly-lensed background
galaxies.

\section{Surface brightness profiles}
\subsection{S\'{e}rsic profile}
The surface photometry of galaxies is often described by a S\'{e}rsic profile \citep{sersic1968}.

\begin{equation}
I(r)=I_{e}\,{\rm exp}\displaystyle \left\{-b_{n}\left[\left(\frac{r}{r_{e}}\right)^{1/n} -1\right]\right \},
\end{equation}

where $I(r)$ is the intensity, $r$ is the radius from the centre of the galaxy, $r_{e}$ is the scale radius, $I_{e}$ is the intensity at $r_{e}$, $n$ in the exponent is a free parameter widely known as the S\'{e}rsic index and $b_{n}=2n - 0.327$; a coefficient chosen so that $r_{e}$ is the half-light radius defined as the radius which encircles half the light from the galaxy \citep{graham1996}. 

This can also be written in terms of the surface brightness, $\mu(r)$ as
\begin{equation}
\mu(r)=\mu_{e} + c_{n}\displaystyle \left[\left(\frac{r}{r_{e}}\right)^{1/n} -1\right],
\end{equation}

where $\mu_{e}$ is the surface brightness at $r_{e}$ and $c_{n}=2.5\,b_{n}$/ln(10).

The S\'{e}rsic profile is a generalisation of the $n=4$ case first used by \cite{devaucouleurs1948} to describe the light profiles of elliptical galaxies and the bulge components of disc galaxies. Depending on the value of $n$, galaxies can be described as disc-like with an $n=1$ exponential profile or bulge-like with higher n values, where ellipticals are expected to have $n=4$ de Vaucouleurs profiles. BCGs with cD-like morphologies (i.e. those with emission from extended halos) are often best fit by a S\'{e}rsic profile with $n>4$ or in some cases by 2 profiles, one for the central bulge and one for the outer envelope \citep{gonzalez2005}, to account for the excess light above that expected from $n=4$ at large radii. 

\section{Analysis and results}
\subsection{Profile fitting}

Due to the degeneracies between $n$ and $r_{e}$ and the potential for a dependency on the surface brightness limit of the images we implement our own 1-D fitting process for which we have greater control over the input data and parameters compared to using `black box' software such as GALFIT \citep{peng2002}. Having said that we first perform 2-dimensional S\'{e}rsic profile fits using the GALFIT (version 3) software package in order to obtain the position angle and ellipticity, of the BCGs. This software requires reasonable initial input parameters such as position, apparent magnitude and ellipticity all of which are estimated by first running the SExtractor package \citep{ba96} so that the iterative fitting process converges to the correct solution in the shortest possible time.  

To create 1-D profiles for our individual BCGs we use the IRAF STScI package Ellipse to extract isophotal intensities with radius, including the ellipticity and position angle parameters derived from GALFIT, while masking out any nearby galaxies interactively. We then fit the resulting output 1-D surface brightness profiles along the semi-major axis of the BCG with both general S\'{e}rsic and de Vaucouleur profiles using a least squares fitting routine. To remove the effect of the PSF we only commence fitting beyond $3\sigma$ of the HST PSF (with the average PSF FWHM for our data $\sim0.11$ arcseconds), however we confirm that the results are robust to beginning this fit further out ($>5\sigma$). These fits are performed to a consistent surface brightness limit $\mu_{\rm F850LP}=27.0$ at $z=1$ which corresponds to $\mu_{\rm F606W}=25.6$ at $z=0.23$ when taking into account surface brightness dimming, a small k correction due to the slight mismatch between the `rest-frame' filters and a correction for passive evolution using a \cite{bc03} simple stellar population (SSP) model with $z_{f}=3$ and solar metallicity, which is a good approximation to the stellar component of these galaxies \citep{collins09,stott2010}. The resulting individual surface brightness profiles for both the low and high redshift samples are found in the Appendix and the individual $r_{e}$ values for the general S\'{e}rsic and de Vaucouleur fits are given in Tables \ref{tab:sample} and \ref{tab:sampleloz}. We note that these 1-D fits are in agreement with the less carefully measured initial output from the GALFIT 2-D fitting software. As with previous studies we find that S\'{e}rsic index and $r_{e}$ are positively correlated (e.g. \citealt{graham1996}). Fitting a surface brightness profile for a BCG can be very difficult particularly when a dominant cD halo component is present as it can sometimes cause the best S\'{e}rsic fit to tend towards $n\rightarrow\infty$, which in practice means $n>>10$. As mentioned in the introduction, the surface brightness profile will then resemble a power law of index $\sim -2$ as intensity falls with radius with the total integrated light diverging, resulting in very large values for $r_{e}$, as seen by others (e.g. \citealt{gonzalez2005}) and here for a number of the low-$z$ clusters. The average semi-major axis scale sizes calculated with a biweight estimator appropriate for non-Gaussian distributions \citep{beers1990} are as follows: for the high redshift BCGs $r_{e}=26.9\pm2.3\rm \,kpc$ ($n=4$) and $r_{e}=57.3\pm15.7\rm\,kpc$, $n=4.3\pm0.9$ (S\'{e}rsic); and for the low redshift sample $r_{e}=43.8\pm5.4\rm \,kpc$ ($n=4$) and $r_{e}=45.5\pm6.9\rm\, kpc$, $n=4.0\pm0.4$ (S\'{e}rsic). 

\begin{figure}
   \centering
     \includegraphics[scale=0.5]{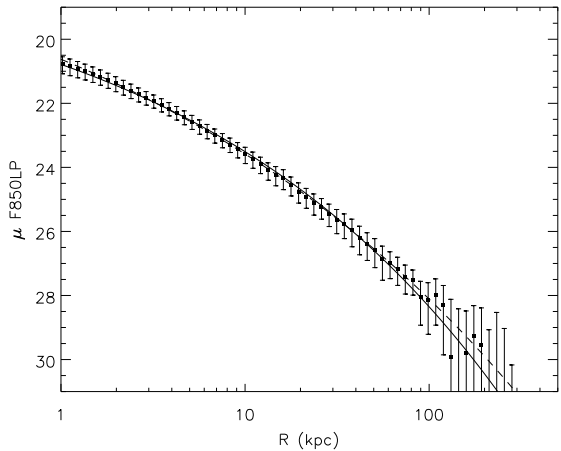} 

     \caption{Surface Brightness profile stack for the high redshift sample with the de Vaucouleur and S\'{e}rsic profile fits plotted (solid and dashed lines respectively).}
   \label{fig:hizstack}
\end{figure}

\begin{figure}
   \centering
     \includegraphics[scale=0.5]{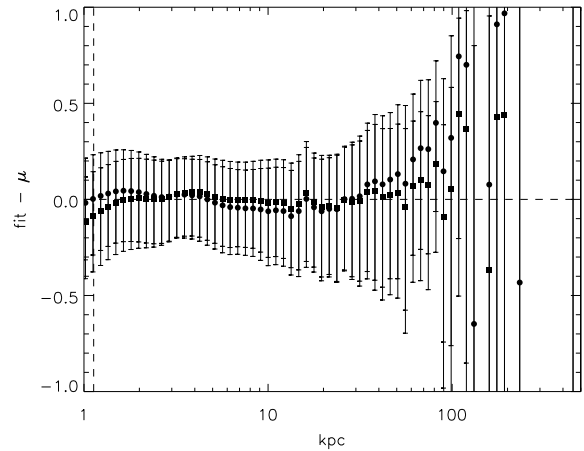} 
     \caption{The difference between the observed surface brightness profile and both the de Vaucouleur and S\'{e}rsic fits for the high redshift stack (circles and squares respectively). The vertical dashed line represents $3\sigma$ of the HST PSF beyond which the fitting commences.}
   \label{fig:hizmod}
\end{figure}

\begin{figure}
   \centering
   \includegraphics[scale=0.5]{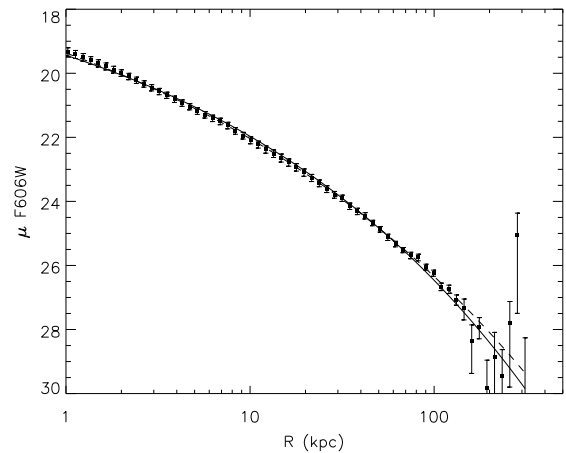}
   \caption{Surface brightness profile stack for the LoCuSS sample with the de Vaucouleur and S\'{e}rsic profile fits plotted (solid and dashed lines respectively).}
   \label{fig:locussstack}
\end{figure}

\begin{figure}
   \centering
     \includegraphics[scale=0.5]{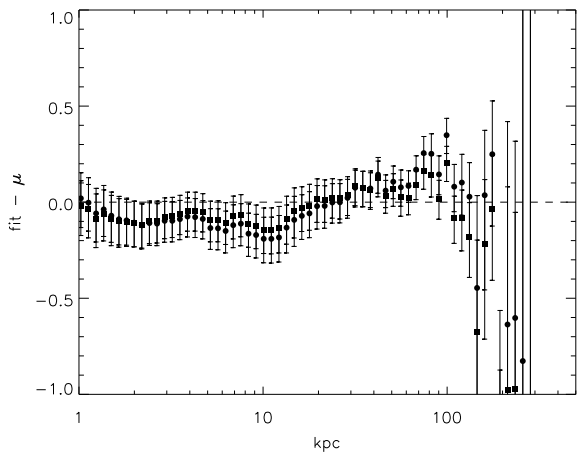} 
     \caption{The difference between the observed surface brightness profile and both the de Vaucouleur and S\'{e}rsic fits for the LoCuSS stack (circles and squares respectively).}
   \label{fig:locussmod}
\end{figure}

\begin{figure}
   \centering
  \includegraphics[scale=0.5]{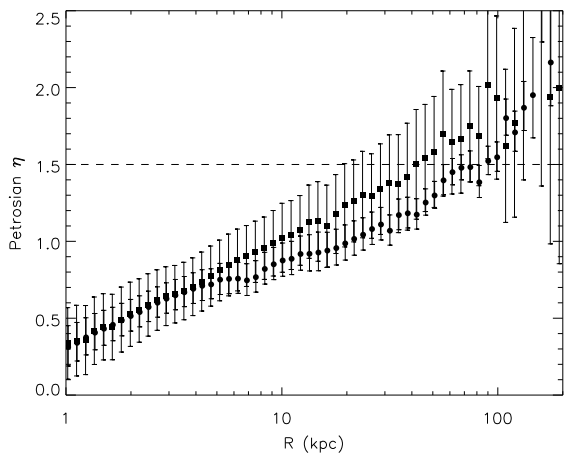} 
   \caption{Petrosian $\eta$ dependency with radius for the low and high redshift stacks (circle and square points respectively). The horizontal line is for $\eta=1.5$ where we measure the Petrosian radius. }
   \label{fig:petro}
\end{figure}

To get a robust measurement of the typical size of a galaxy at each epoch we also stack our low and high redshift 1-D surface brightness profiles in order to provide a smoother and deeper dataset. We do this first for the high redshift sample, correcting all of the surface brightness profiles to $z=1$, accounting for the effect of cosmic dimming, angular scale size and both k and evolution corrections based on a \cite{bc03} SSP model with solar metallicity and a formation redshift $z_{f}=3$ as above. The same is then done for the low redshift clusters correcting to $z=0.23$, the average redshift of the sample. We then fit the high and low redshift surface brightness stacks, again beyond $3\sigma$ of the HST PSF, using the same 1-D fitting code and find that for the high redshift BCGs: $r_{e}=32.1\pm2.5\rm\, kpc$ when $n=4$ and $r_{e}=47.6\pm13.7\rm\, kpc$ with $n=5.4\pm0.9$ for the best fit S\'{e}rsic model; while for the low redshift sample: $r_{e}=43.2\pm1.0\rm\, kpc$ when $n=4$ and for a free S\'{e}rsic fit $r_{e}=57.9\pm4.5\rm\, kpc$ with $n=4.8\pm0.2$, in good agreement with the average of the individual 1-D fits. These fits are performed to a deeper surface brightness limit, $\mu_{\rm F850LP}=28.0$ at $z=1$ corresponding to $\mu_{\rm F606W}=26.6$ at $z=0.23$ and we note that if the fits are instead performed to the original shallower limits of the individual fits the results are still in agreement. The resulting stacked surface brightness profiles and their de Vaucouleur and S\'{e}rsic fits are shown in Figures \ref{fig:hizstack} and \ref{fig:locussstack} with the corresponding residuals about these fits shown in Figures \ref{fig:hizmod} and \ref{fig:locussmod}. We also look at the Petrosian radius  \citep{petrosian1976}, which has the advantage of not being so affected by the background and photometric uncertainties as $r_e$. Fixing $\eta=1.5$ for comparison with \cite{brough2005}, gives $r_{pet}=41.9\rm\, kpc$ and $r_{pet}=61.7\rm\, kpc$ for the high and low redshift stacks respectively (see Figure \ref{fig:petro}). Finally to get a further non-parametric estimate of the scale size we integrate the total light within the stacks down to the same surface brightness limits as the fits and find that the half-light radius $r_{1/2}=23.6\pm9.6\rm\, kpc$ for the high z data and $r_{1/2}=27.0\pm3.3\rm\, kpc$ for the low-$z$ sample.

The results of the average scale size, stack fitting and non-parametric analysis are found in Table \ref{tab:size}. From these results we can see that there is evidence for at most a weak trend in scale size or profile shape for BCGs over a period of 5\,Gyr, between $z=1$ and $z=0.2$. If we consider just the de Vaucouleur fits to the stacked clusters and the average of the individual de Vaucouleur fits, then the high redshift galaxies at $z=1$ are still only found to be a factor $0.6 - 0.7$ the size of their $z=0.2$ counterparts. This is the maximum evolution implied by our data as the S\'{e}rsic fits and
integrated light size estimator suggest this is considerably less. Clearly there is no evidence for the large growth factors of order 4 reported for other early-type galaxies over a similar epochs. In addition to this 3 of the individual S\'{e}rsic fits at high redshift show significant cD halos as their $n$ values are significantly greater than the $n=4$ de Vaucouleur prediction appropriate to a regular elliptical galaxy.

%As a check on our previous results that there has been little stellar mass evolution of BCGs over a similar time period \citep{collins09,stott2010}, we note that the observed magnitudes enclosed by a 25 kpc radius aperture for the high and low redshift stacks are entirely consistent with there being only passive evolution during the intervening period when we compare the values. Using a \cite{bc03} model with $z_{f}=3$ and solar metallicity evolved from the observed magnitude at $z=1$ (namely $M_{F850LP}=20.8\pm0.3$) we predict the corresponding value at z=0.23 to be $M_{F606W}=17.7\pm0.3$, which compares well with the measured value of $M_{F606W}=17.4\pm0.1$. When measuring the flux instead to a consistent surface brightness limit , $\mu_{F850LP}=25.0$ at $z=1$ ($\mu_{F606W}=23.6$ at z=0.23), we find $M_{F850LP}=20.6\pm0.1$ at $z=1$ predicted to be $M_{F606W}=17.5\pm0.1$ at $z=0.23$ and measure $M_{F606W}=17.4\pm0.1$ which is again fully consistent with passive evolution. This is of course only a crude proxy for the stellar mass and is for a part of the galaxy spectrum that is too blue to be used for stellar mass estimates but it is reassuring to see a consistent picture emerge.

\begin{table*}
\begin{center}
\caption[]{Summary of measured parameters. $r_{pet}$ is the Petrosian radius where the Petrosian parameter $\eta=1.5$. $r_{1/2}$ is a non-parametric half-light radius derived from the integrated flux.}
\label{tab:size}
\small\begin{tabular}{lllllllll}

\hline
&\multicolumn{3}{c}{average}&\multicolumn{5}{c}{stack}\\
$z$&$r_{e} (n=4)$&$r_{e}$ (free)&n&$r_{e} (n=4)$&$r_{e}$ (free)&n&$r_{pet}$&$r_{1/2}$\\
&(kpc)&(kpc)&& (kpc)&(kpc)&&(kpc)&(kpc)\\
\hline
0.23&$43.8\pm5.4$&$45.5\pm6.9$&$4.0\pm0.4$&$43.2\pm1.0$&$57.9\pm4.5$&$4.8\pm0.2$&$61.7$&$27.0\pm3.3$\\
1.00&$26.9\pm2.3$&$57.3\pm15.7$&$4.3\pm0.9$&$32.1\pm2.5$&$47.6\pm13.7$&$5.4\pm0.9$&$41.9$&$23.6\pm9.6$\\

\hline
\end{tabular}
\end{center}
\end{table*}

\subsection{Kormendy relation}
\begin{figure}
   \centering
    \includegraphics[scale=0.5]{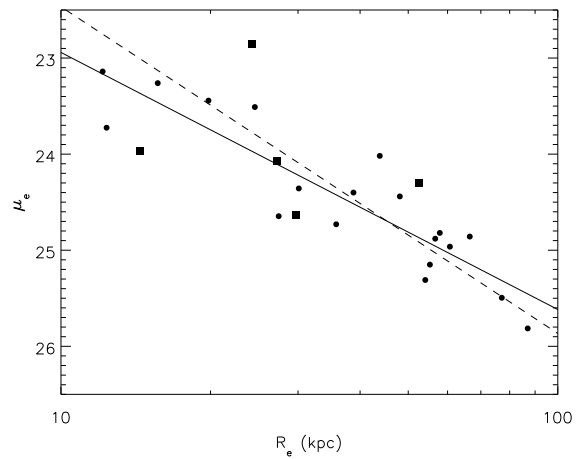}
   \caption{The Kormendy relation for our sample. The LoCuSS and the high redshift BCGs (accounting for cosmological dimming and k and evolution correction to match the LoCuSS sample) are represented by circles and squares respectively.}
   \label{fig:korm}
\end{figure}

The Kormendy relation is a slice through the fundamental plane that shows how the scale size of galaxies depends on the surface brightness \citep{kormendy1977}. Formally it is a plot of the surface brightness at the effective radius ($\mu_e$) vs $r_{e}$ for a de Vaucouleur surface brightness fit. BCGs are found to lie off the relation for normal elliptical galaxies as they possess large extended halos  \citep{hoessel1987,schombert1987, oegerle1991}. A similar relation to this is BCG size versus luminosity, which for BCGs is found to be steeper than that of regular elliptical galaxies \citep{bernardi2007,liu2008}, although, as mentioned in the introduction, this is a matter for some debate as authors using essentially the same data find a significantly shallower slope \citep{vonderlinden2007}. In Figure \ref{fig:korm} we plot the Kormendy relation for our samples. The slope of this relation is parameterised as $\mu_e=A+B\,\rm log \,r_{e}$ and for the LoCuSS sample $A=20.3\pm0.9$ and $B=2.7\pm0.8$, with the errors reflecting the intrinsic scatter in the relation, not just the formal error of the fit. This is in good agreement with the work of \cite{brough2005} who find $B=2.60\pm0.03$ but we cannot rule out larger values of $B$ found in other studies, e.g. $B=3.1\pm0.1$ found by both \cite{hoessel1987} and \cite{oegerle1991} and $B=3.44\pm0.13$ found by \cite{bildfell2008}. We include our fit (solid line) and a dashed line with the same slope as that found by \cite{bildfell2008} for comparison. It is obviously not possible to perform a fit to the high redshift BCGs and look for an evolution in the Kormendy relation with only a handful of points but we include them on the plot after correcting for cosmic dimming and k and evolution correction using a  \cite{bc03} SSP model with solar metallicity and a formation redshift $z_{f}=3$ as above. As expected from the surface brightness fitting results, the high redshift points are found to occupy the same region of the plot as the low redshift relation.  

\section[]{Discussion}
The results of our investigation demonstrate that when the best available imaging data are used there is evidence for at most only a small increase in the scale size of BCGs over the 5\,Gyrs between $z=1$ and $z=0.2$ and no evidence for a change in the shape of their light profiles. This is in contrast with other BCG studies that probe the scale size to moderate redshifts (e.g. \citealt{nelson2002b,ascaso2010}) and with the results for massive field galaxies (e.g. \citealt{vandokkum2009,vandokkum2010}). The interpretation of this result when taken in concert with the findings of our near-infrared observations, which demonstrate that there has been little evolution in the stellar mass over the same period, suggests that major merging is unlikely to be an important process for BCG evolution since $z=1$ \citep{collins09,stott2010}. This contrasts with semi-analytic models of BCG evolution based on hierarchical N-body simulations which require $\sim70\%$ of the final BCG stellar mass to be accreted over this time frame \citep{deluc07}. Other models have tried to predict the scale size evolution for massive elliptical galaxies and also find significant growth factors since redshift 1 (e.g. \citealt{kochfar2006}; \citealt{naab09}). 

As discussed in the introduction, a detailed comparison with other results is difficult due to the heterogeneity of the samples and analyses, so here we restrict  ourselves to comparing general trends.  Our results do not support the findings of \cite{valentinuzzi2010b} who report an average  $r_e$ value 
of only $8\,$kpc  using 8 clusters from the ESO Distant Clusters Survey at $0.4<z<0.8$, however this sample is not X-ray selected and may contain BCGs which do not trace the more massive cD systems associated with X-ray samples at high redshift \citep{bcm00}. 

There are two other recent results reporting evolution of BCG size at intermediate redshift: first, \citealt{bernardi2009}, find that BCGs in SDSS optically selected clusters at $z\simeq 0.25$ are smaller than their lower redshift counterparts by as much as $70\%$; secondly,  \cite{ascaso2010} find a reduction of 0.6 in relative scale size between 20 BCGs in the range $0.3<z<0.6$)  and an X-ray matched sample at $z\sim0.05$. This would be consistent with our result if the evolution takes place relatively recently ($z<0.25$), as suggested by \cite{bernardi2009}. Having said that we do find on average larger BCG $r_{e}$ values at $z=0.2$ than in these studies and the average $r_e$ ($44\pm 5\rm\,kpc$) for our LoCuSS sample is more consistent with the average $r_e$ value ($52\pm 4\rm\,kpc$) found by \cite{bildfell2008} for 48 X-ray luminous clusters at an average redshift of $0.26$ and is arguably a better sample to compare with LoCuSS.

Restricting our analysis to only the highest quality data available for the highest redshift X-ray clusters, we have found that no great size evolution exists for BCGs. As far as the result for the field galaxies goes we can only speculate that either there is a strong environmental dependence, with massive galaxies in clusters already being morphologically mature at high redshift whereas massive field galaxies are yet to undergo a transformation. On the other hand, as we emphasised in the introduction, it is not possible to rule out observational bias.

We note that the extended cD halo is often found to modify the BCG profile to have a higher S\'{e}rsic index than the standard elliptical-like de Vaucouleur $n=4$. As 3 of the individual S\'{e}rsic fits at high redshift have $n>4$, we conclude that the cD halo is in place for at least some BCGs at high redshift, although these are washed out to some extent in the stack. However, some models predict that this halo forms late and results in BCGs only recently departing from the Kormendy relation for normal ellipticals \citep{rands09}. It would be interesting to obtain deep data in other optical bands to see whether this cD halo has a significant colour gradient at high redshift compared to local samples as then we may begin to investigate its age and origin if significantly different in colour to the main bulge of the BCG. 

Due to our clusters being high redshift X-ray emitters discovered in flux-limited X-ray surveys, there is an argument that they represent the most massive and therefore rarest objects at that epoch and therefore we may not be making a fair comparison. One may expect that these massive clusters would themselves contain the most massive and therefore by inference the largest sized BCGs at that epoch. However, when we look at our low redshift sample we find no correlation between X-ray luminosity or temperature and $r_{e}$ in agreement with \citealt{ascaso2010} and so we conclude this is not a significant issue. One other aspect is the affect that cluster cool-core strength has on the properties of BCGs, particularly as the number of cool-core clusters appears to evolve \citep{santos2010}, possibly resulting in inconsistent X-ray selection with redshift. However it is reassuring that a quick check using 18 of the low-$z$ clusters observed with Chandra, shows no correlation between the inner gas density profile slope and the BCG scale size (A. Sanderson, private communication). Another possible bias is that to be discovered by X-ray surveys at such high redshifts, we are only seeing the most relaxed clusters with the highest central gas densities which may be expected to host more morphologically mature BCGs at that epoch. Of course we cannot account for this and other biases with such a small sample selected from an unknown mass function at high redshift. A larger sample of BCGs is required above $z=1$ , including those hosted by less relaxed cluster systems, to account for potential biases.

Our evidence supports assertion that BCGs do not change in 
appearance over the last 6\,Gyr and occupy a similar part of the Kormendy relation over this entire period. Interestingly the other galaxy type for which the Kormendy relation appears not to have changed over a similar timeframe is radio galaxies. Although these objects have scale sizes less than BCGs ($r_e\simeq 12-15$ kpc at $z\sim0.5-1$ \citep{mclure2004}, only modest size evolution has been seen in powerful radio galaxies, growing from $\sim8$kpc at $z=2$ a factor of 1.5 by the present day \citep{targett2010}. This may be suggestive that radio galaxies 
and BCGs evolve along similar evolutionary paths, at least at late times. 

We have a picture of large BCGs residing in the cores of massive clusters with their cD halos already in place by $z=1$ and any subsequent merger or interaction events having, on average, very little effect on the global properties of the galaxy. This may be understood if we assume that major merger events in the cores of massive clusters at $z=1$ will be just as rare as their local counterparts due to the high velocity dispersions involved. Significant amounts of star formation will also be suppressed by the hot cluster environment so we can perhaps think of these systems as already fully mature at this epoch. The discovery of galaxies with large stellar mass and scale size at these epochs is a major challenge to the current theoretical models, although comparisons using deeper imaging of larger cluster mass-selected samples are required to make 
further definitive tests.

\section*{Acknowledgments}
Firstly, we would like to thank the anonymous referee for their useful comments and suggestions.

We thank Al Sanderson for providing the cool-core strength information mentioned in the discussion.

We acknowledge financial support from Liverpool John Moores University and the STFC. GPS acknowledges support from the Royal Society

IRAF is distributed by the National Optical Astronomy Observatories, which are operated by the Association of Universities for Research in Astronomy, Inc., under cooperative agreement with the National Science Foundation.

\bibliographystyle{mn2e}
\bibliography{bcgsize}

%\newpage
\appendix
\section{Surface brightness profiles}
\begin{figure*}
   \centering
  %remember to copy all into folder for paper when submitting
%\includegraphics[scale=0.5]{/Users/jps/radius/locuss/rxcj0043/sbp.ps} 
%\includegraphics[scale=0.5]{/Users/jps/radius/locuss/rxcj0105/sbp.ps} 
%\includegraphics[scale=0.5]{/Users/jps/radius/locuss/rxcj0118/sbp.ps} 
%\includegraphics[scale=0.5]{/Users/jps/radius/locuss/rxcj0220/sbp.ps} 
%\includegraphics[scale=0.5]{/Users/jps/radius/locuss/rxcj0237/sbp.ps} 
%\includegraphics[scale=0.5]{/Users/jps/radius/locuss/rxcj0304/sbp.ps} 
\includegraphics[scale=0.5]{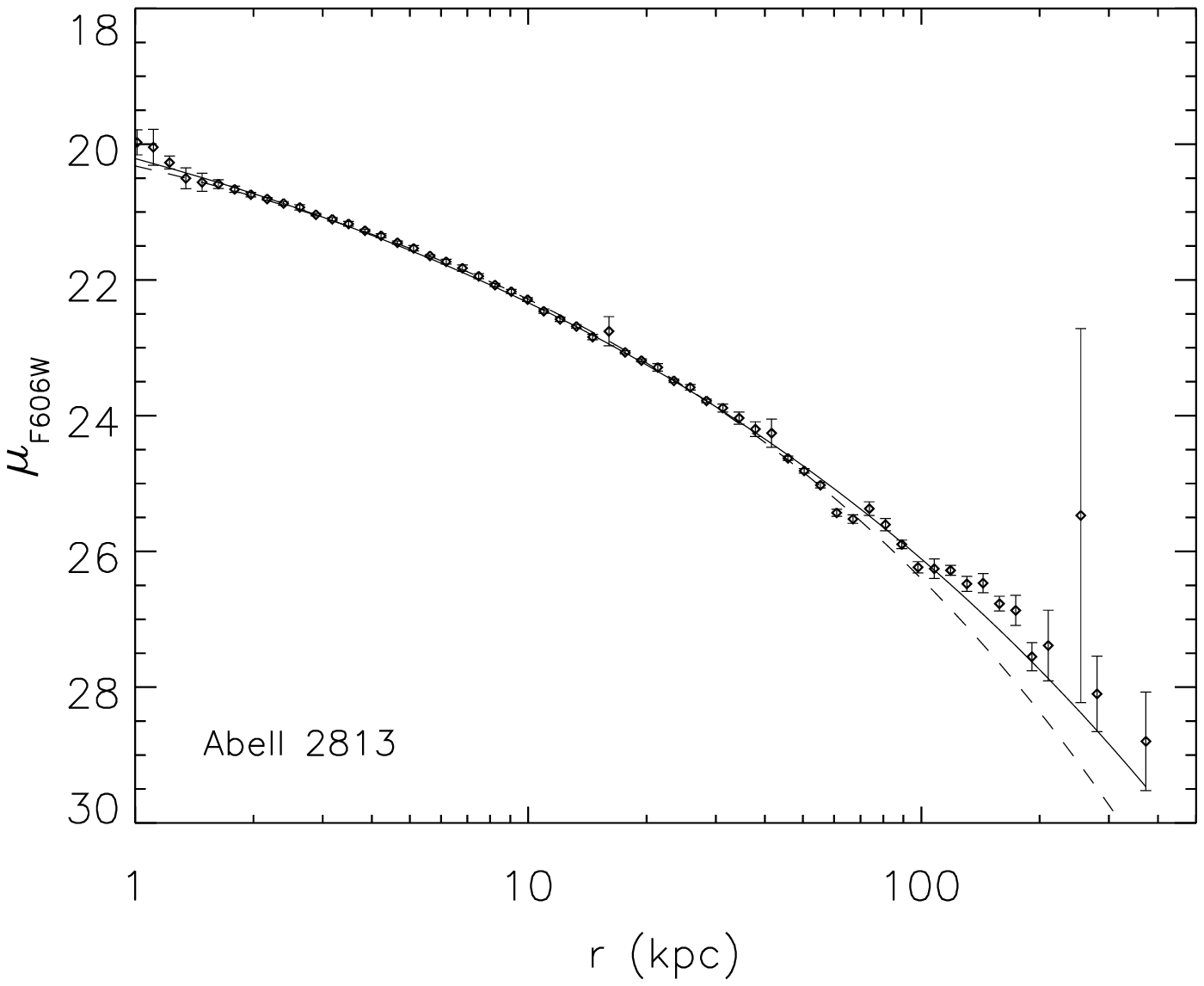} 
\includegraphics[scale=0.5]{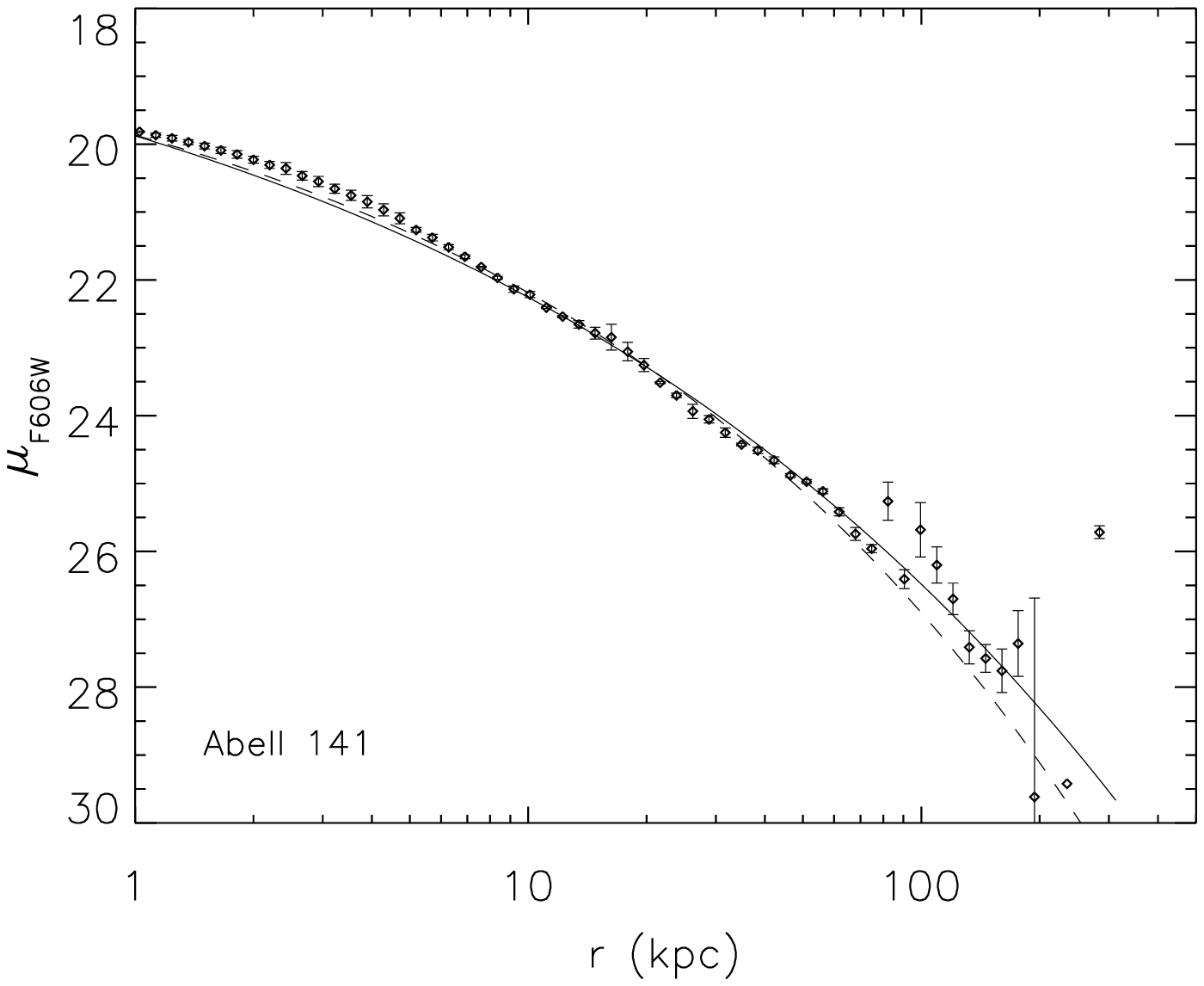} 
\includegraphics[scale=0.5]{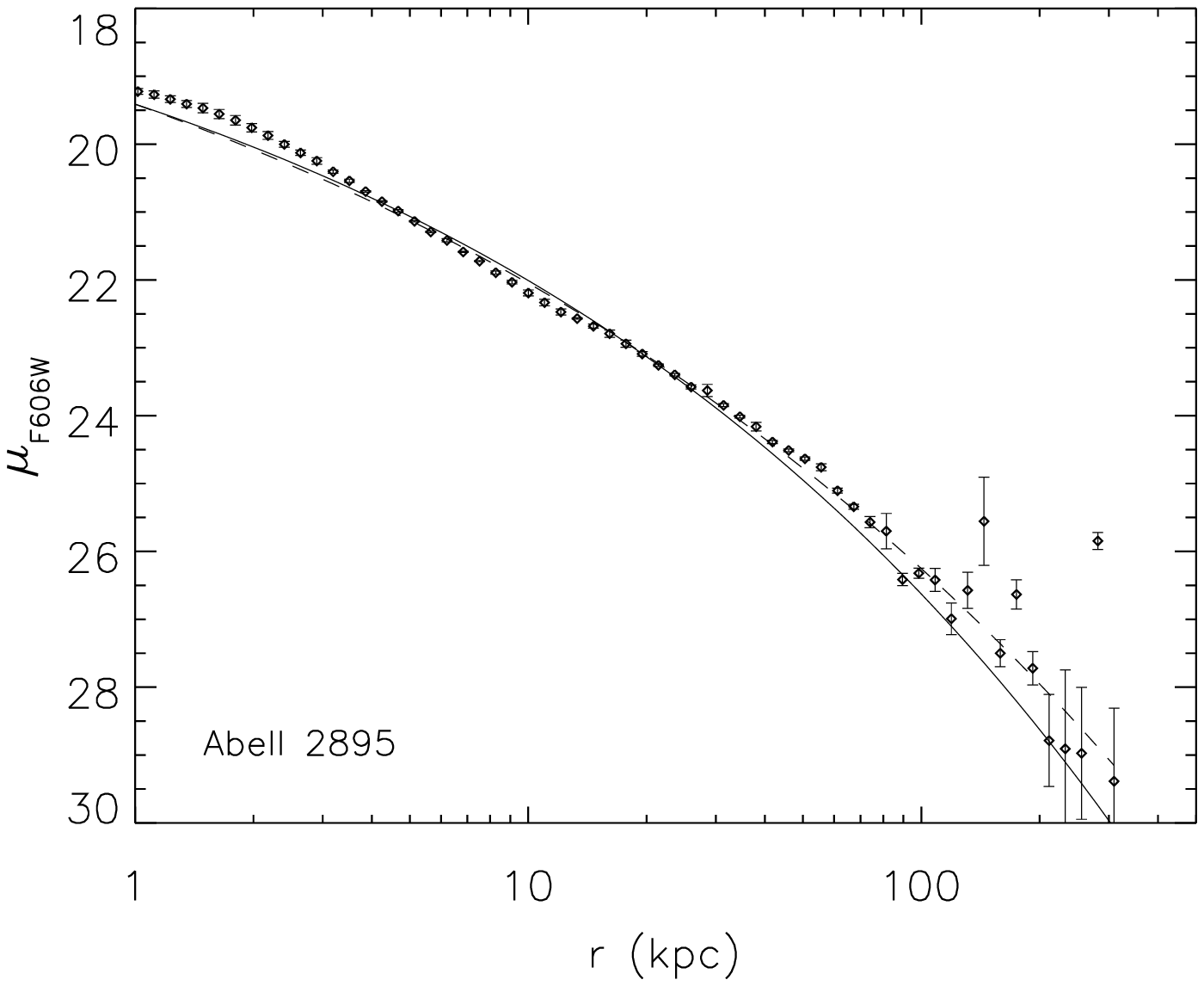} 
\includegraphics[scale=0.5]{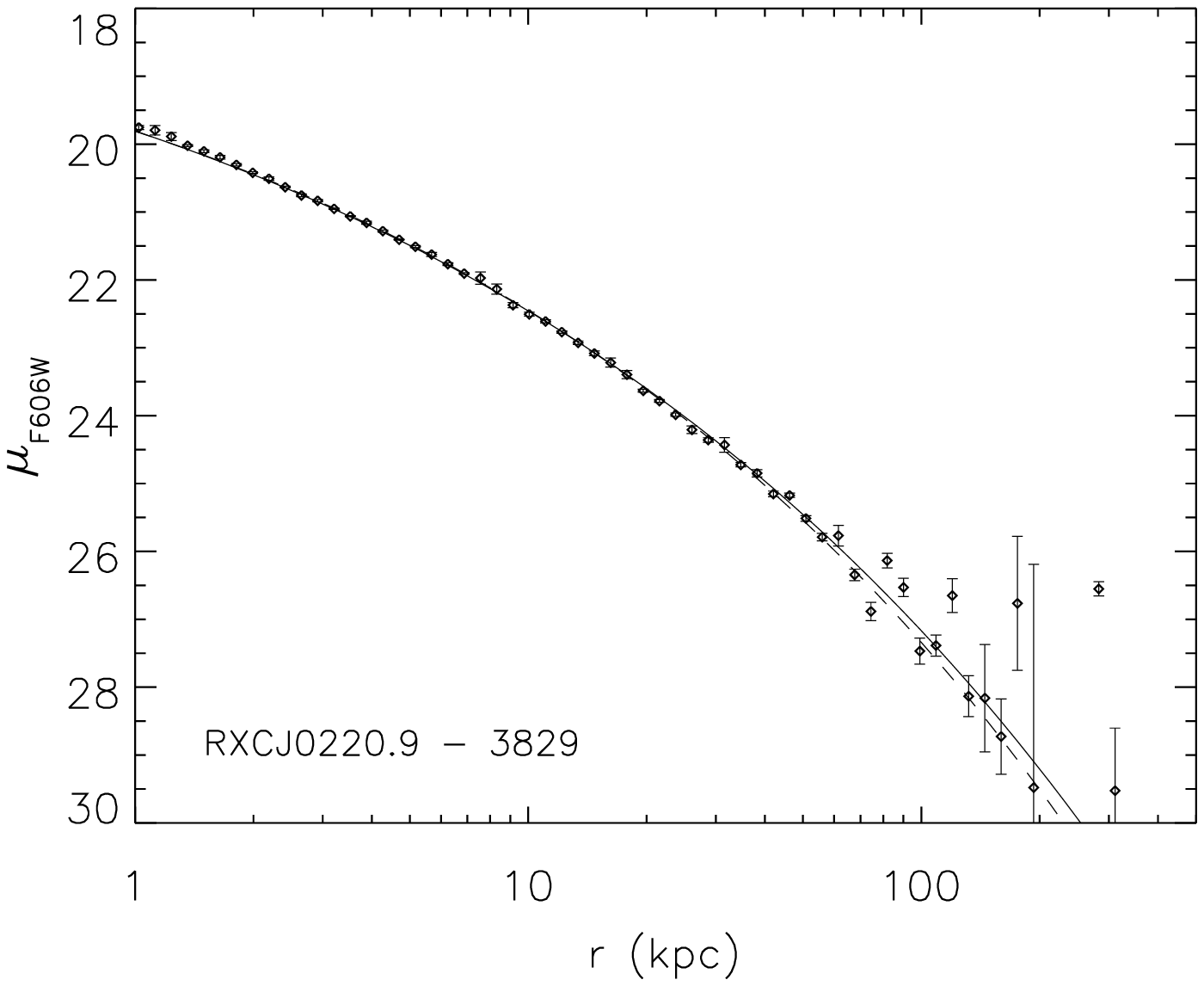} 
\includegraphics[scale=0.5]{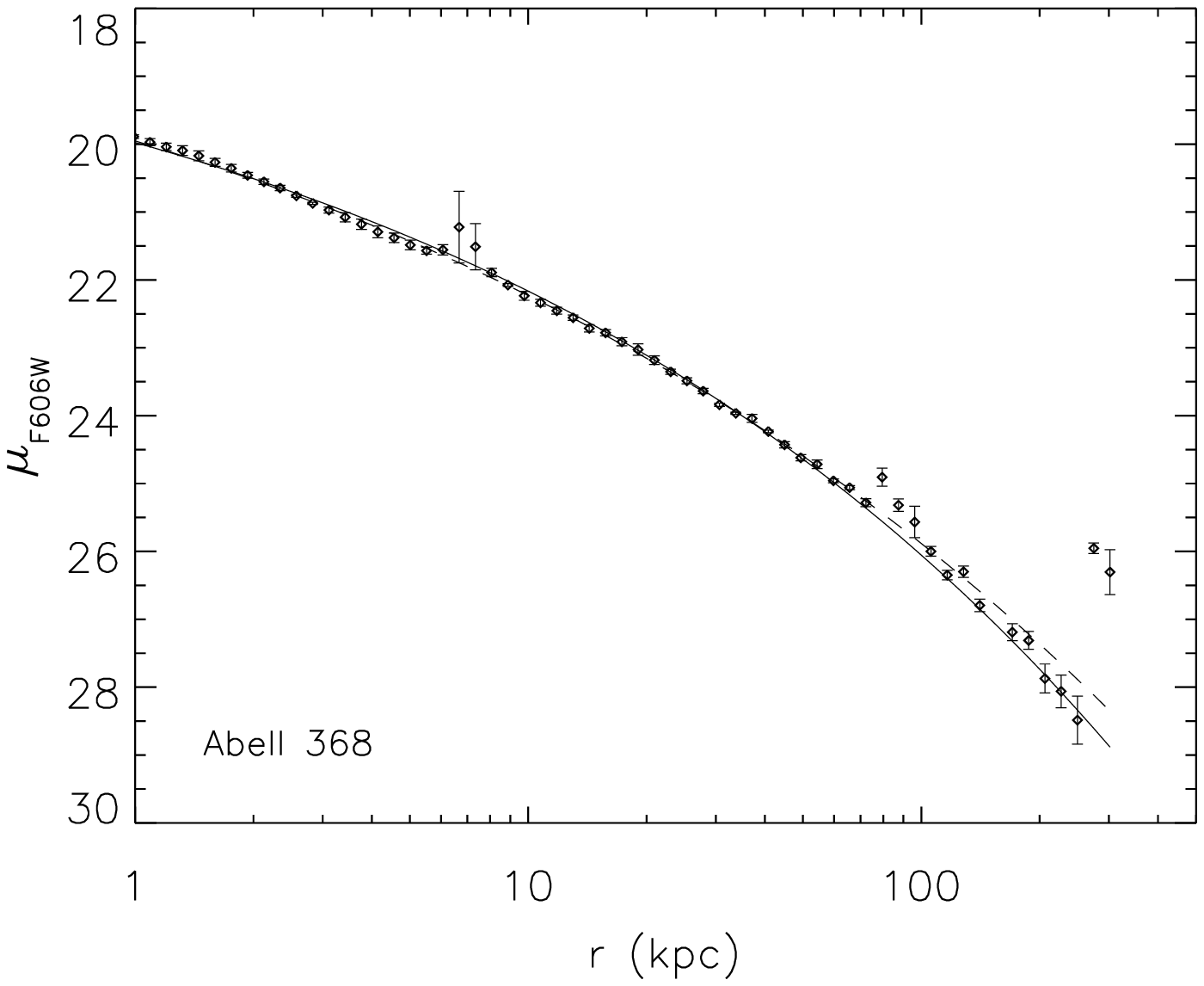} 
\includegraphics[scale=0.5]{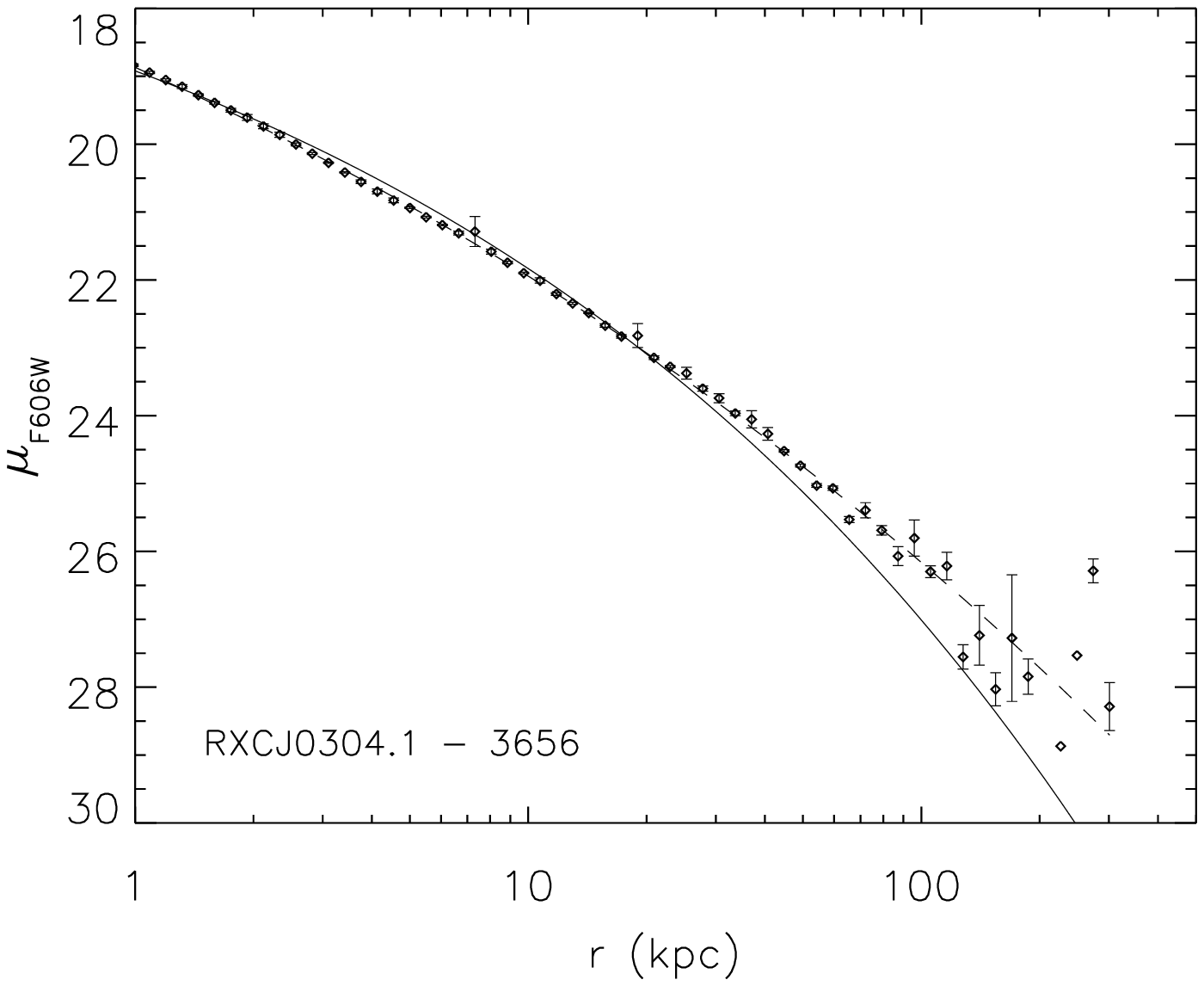} 

\caption{The surface brightness profiles for the individual LoCuSS BCGs 1.  de Vaucouleur and S\'{e}rsic profile fits plotted (solid and dashed lines respectively).}
   \label{fig:locuss1}
\end{figure*}

\begin{figure*}
   \centering
\includegraphics[scale=0.5]{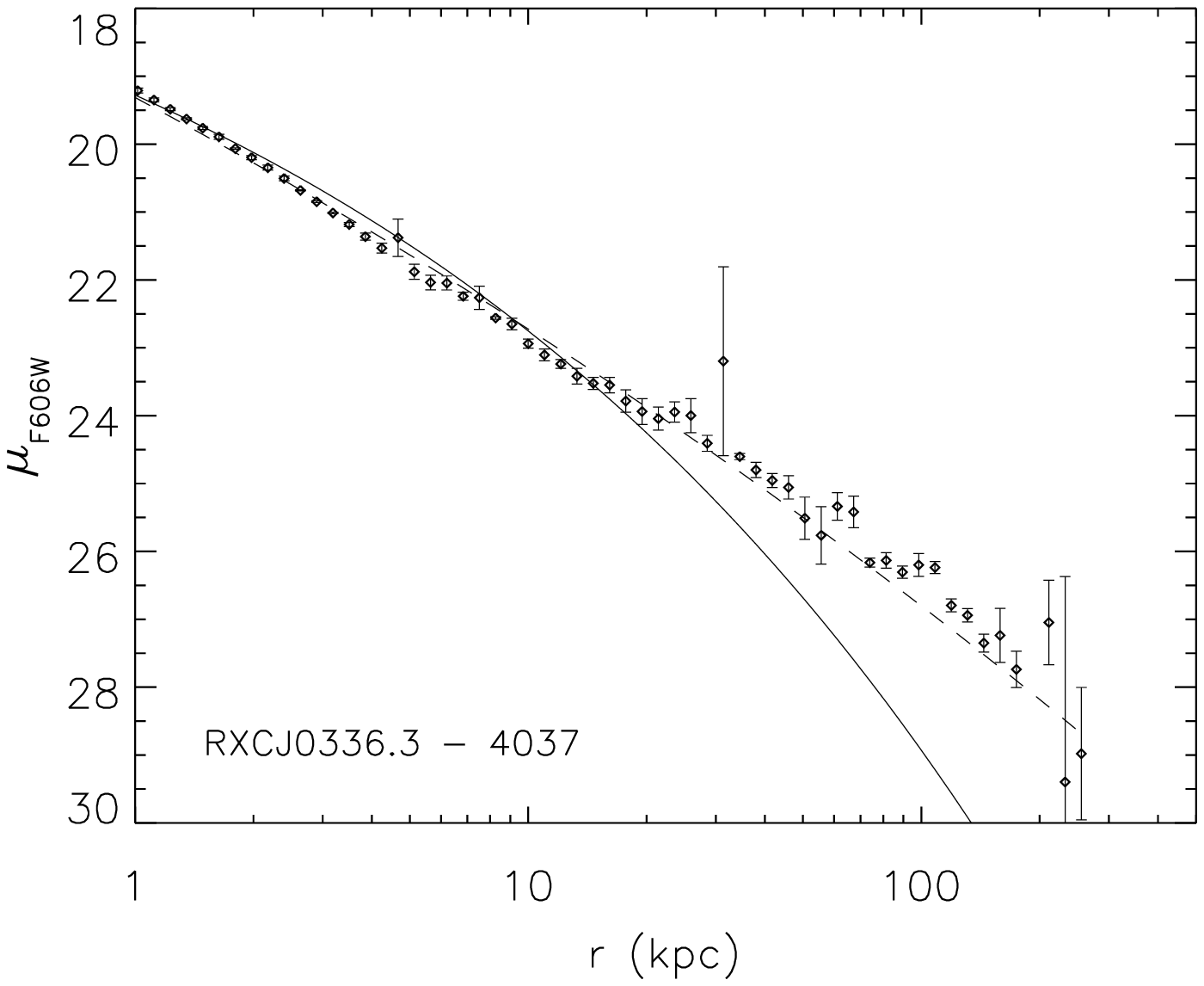} 
\includegraphics[scale=0.5]{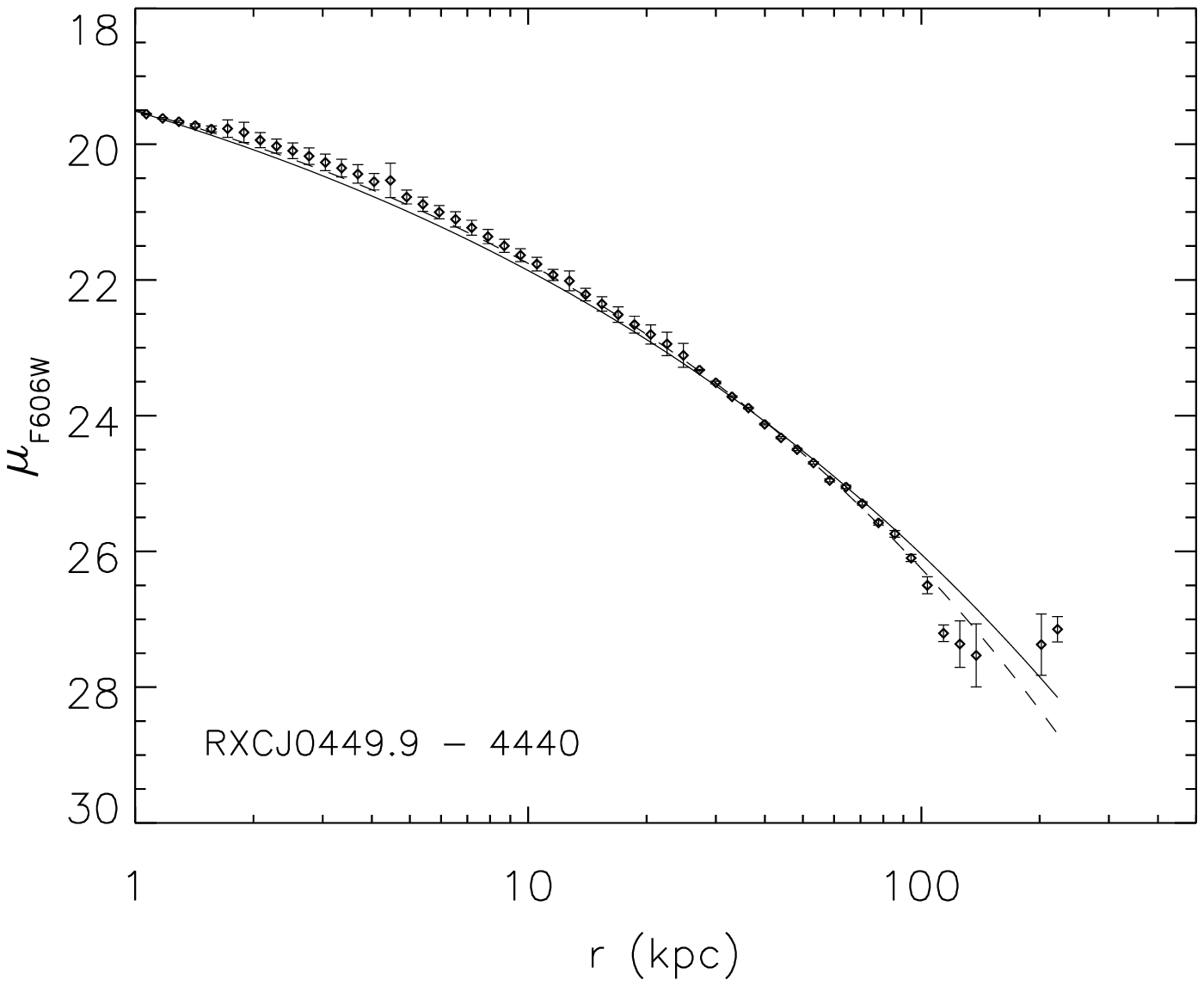} 
\includegraphics[scale=0.5]{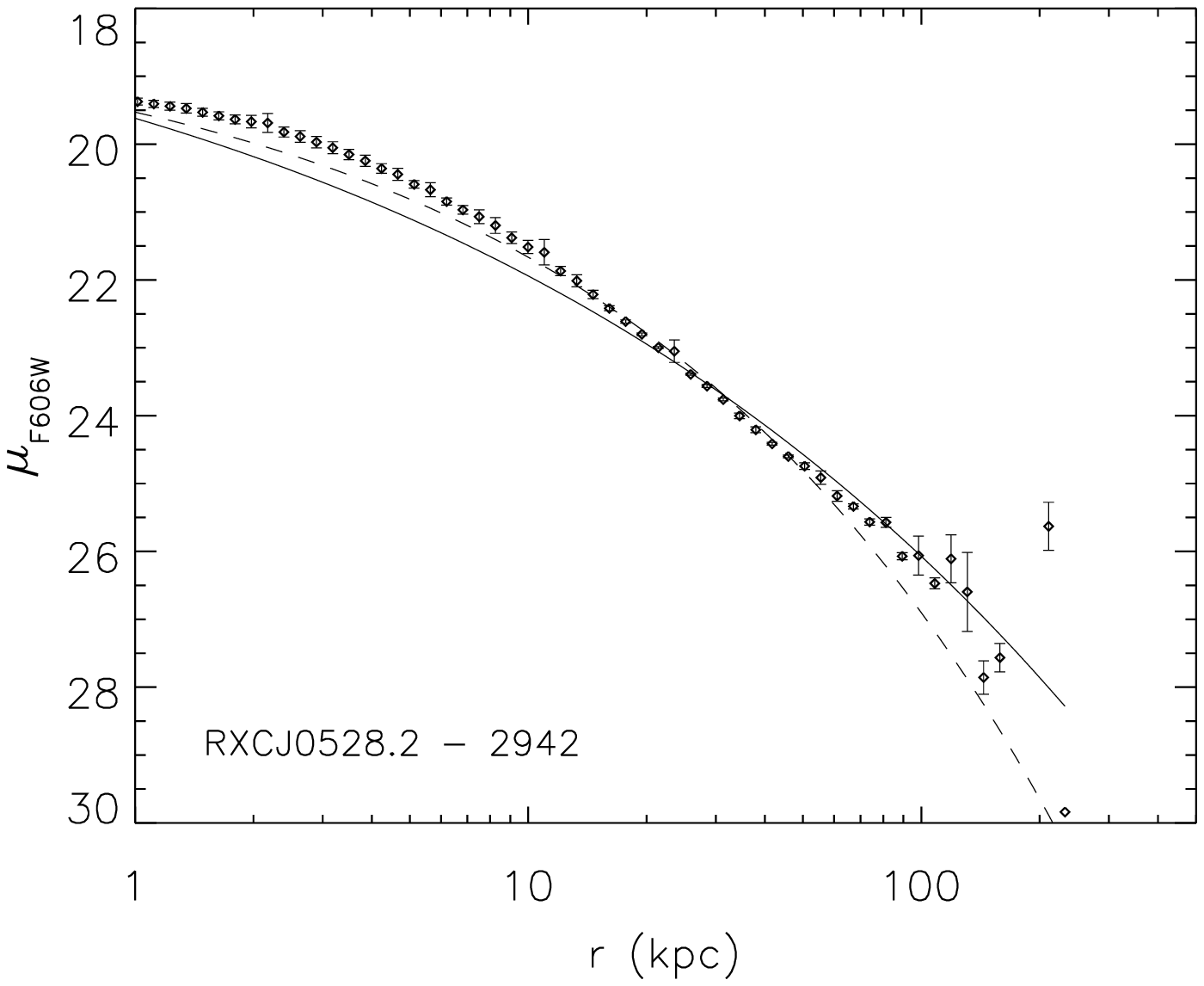} 
\includegraphics[scale=0.5]{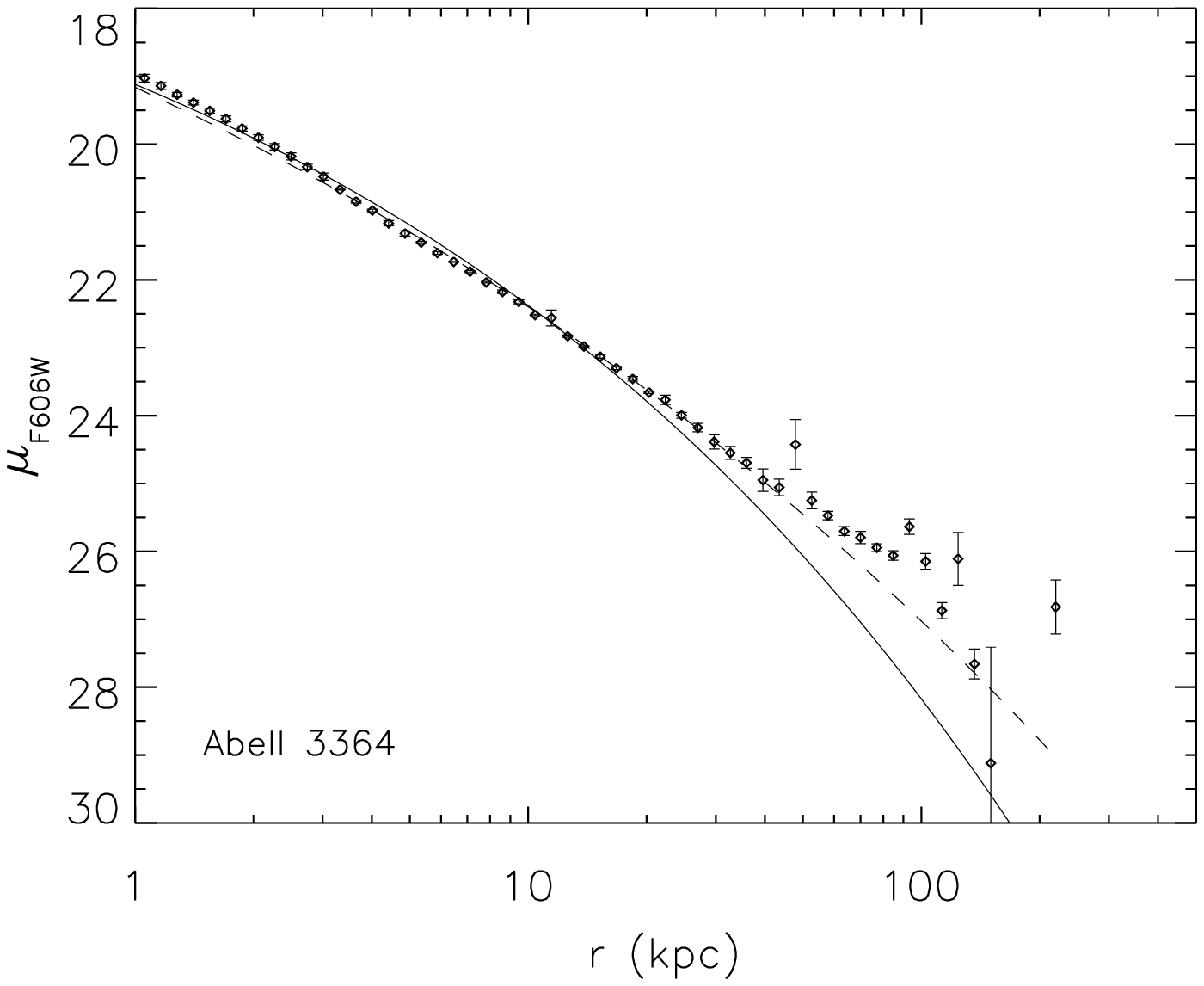} 
\includegraphics[scale=0.5]{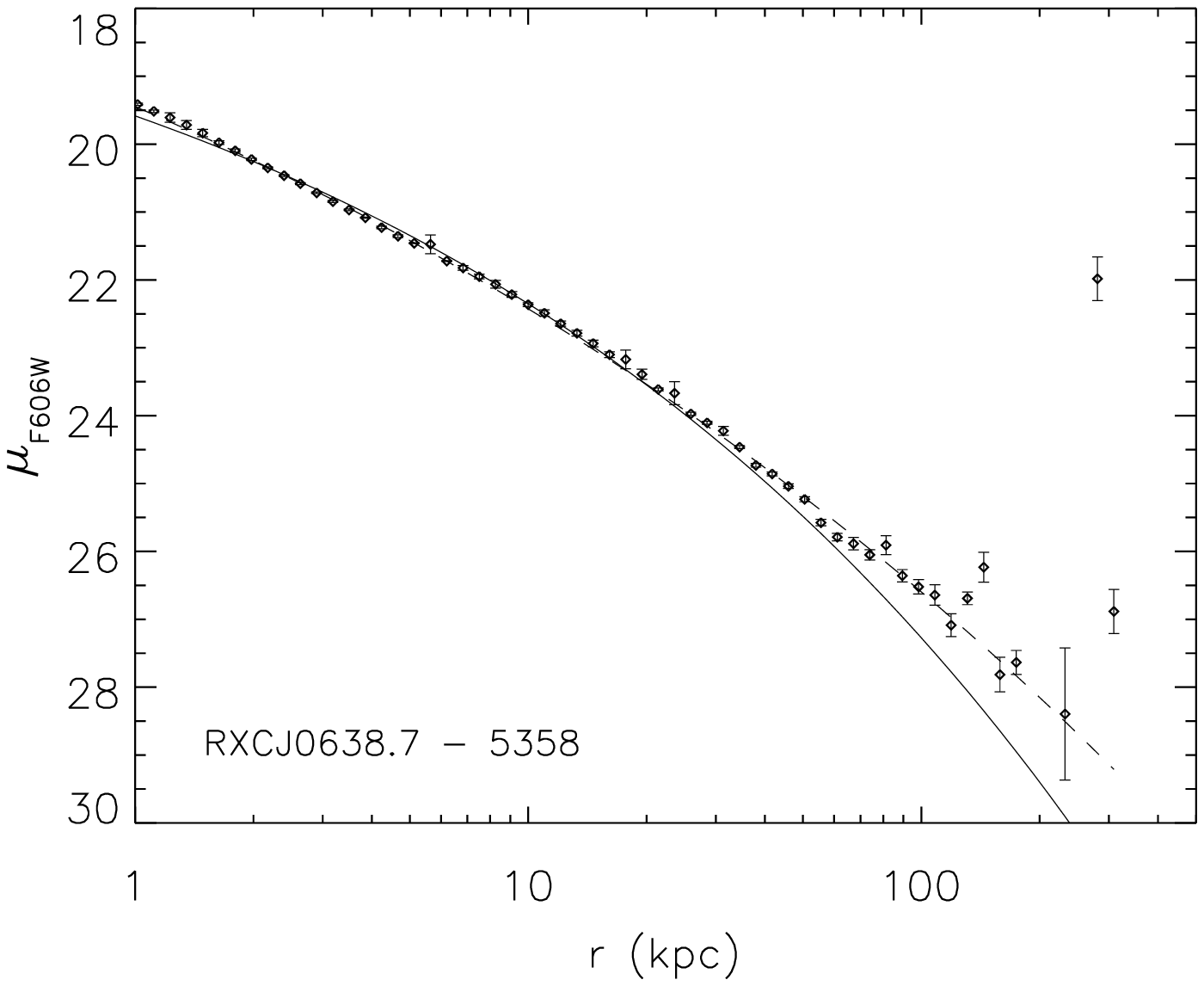} 
\includegraphics[scale=0.5]{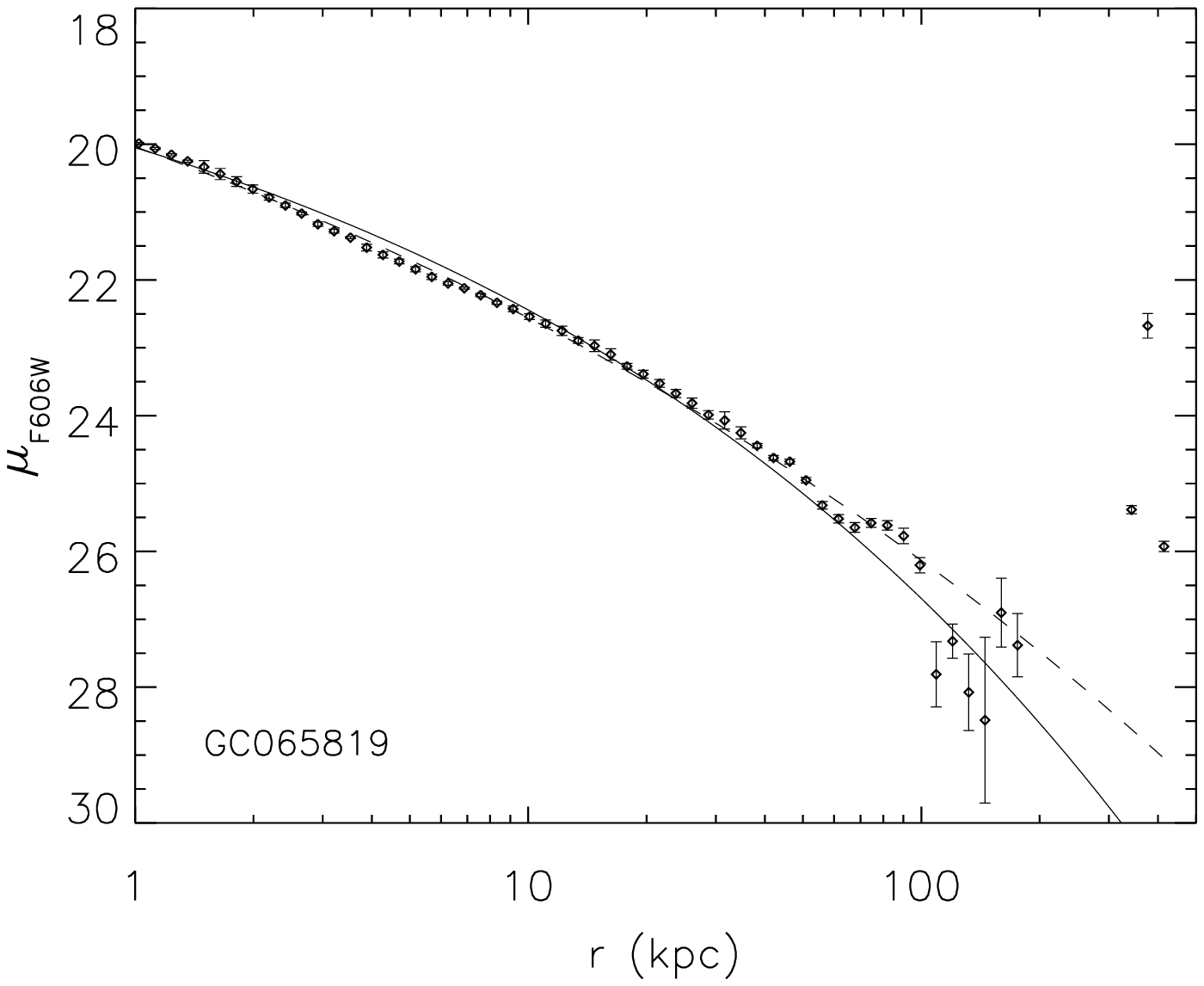} 
\caption{The surface brightness profiles for the individual LoCuSS BCGs 2.  de Vaucouleur and S\'{e}rsic profile fits plotted (solid and dashed lines respectively).}
   \label{fig:locuss2}
\end{figure*}

\begin{figure*}
   \centering
\includegraphics[scale=0.5]{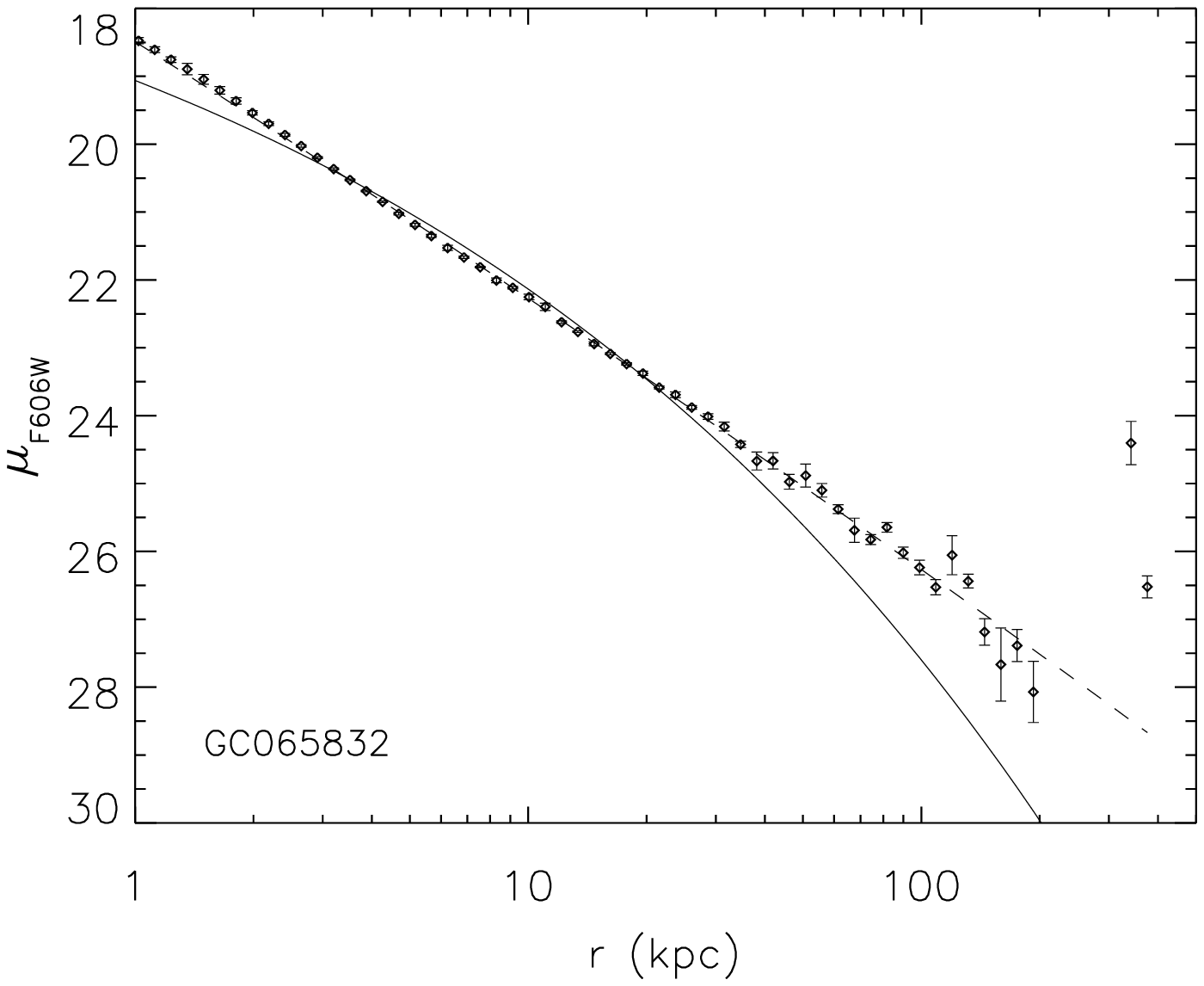} 
\includegraphics[scale=0.5]{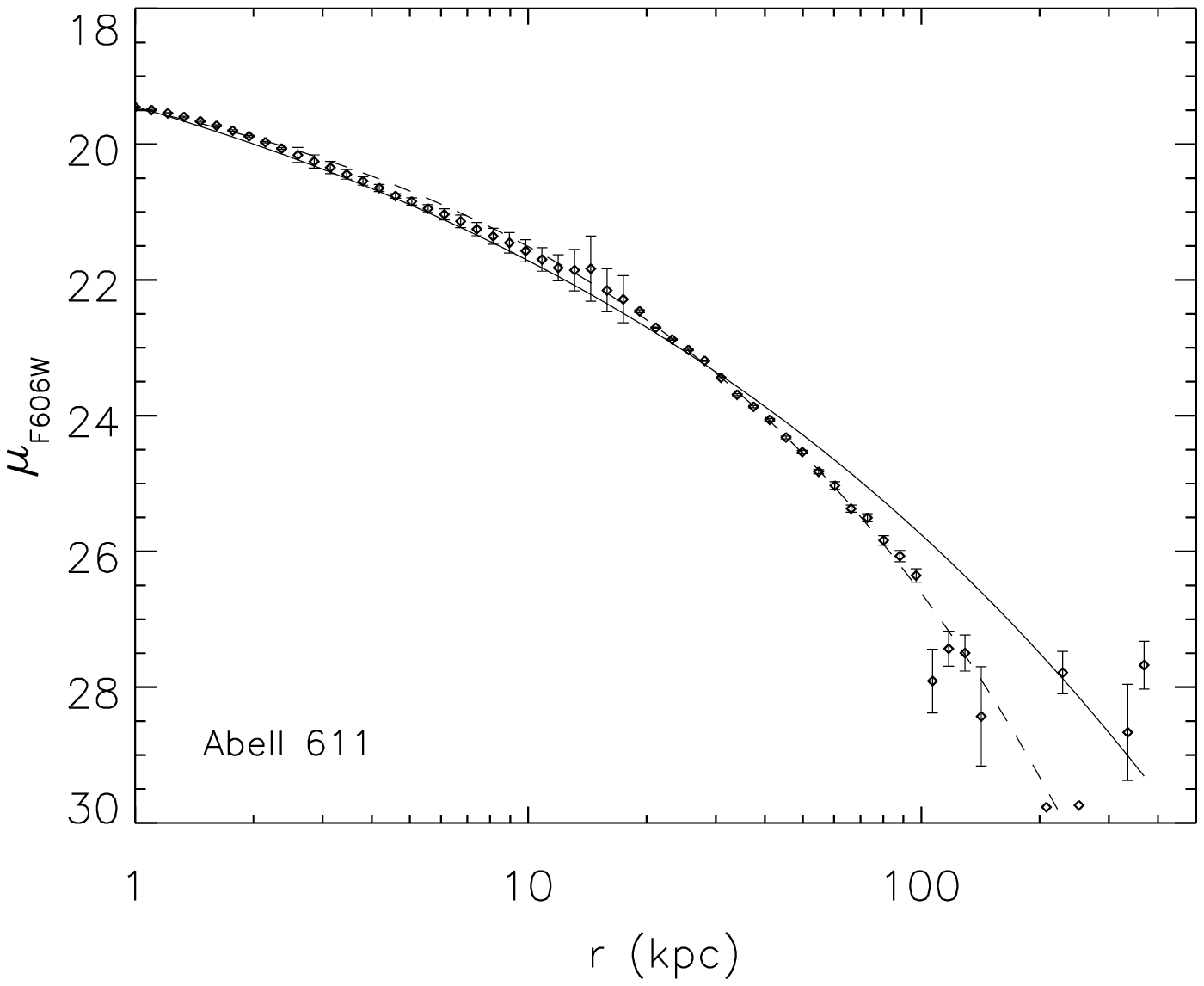} 
\includegraphics[scale=0.5]{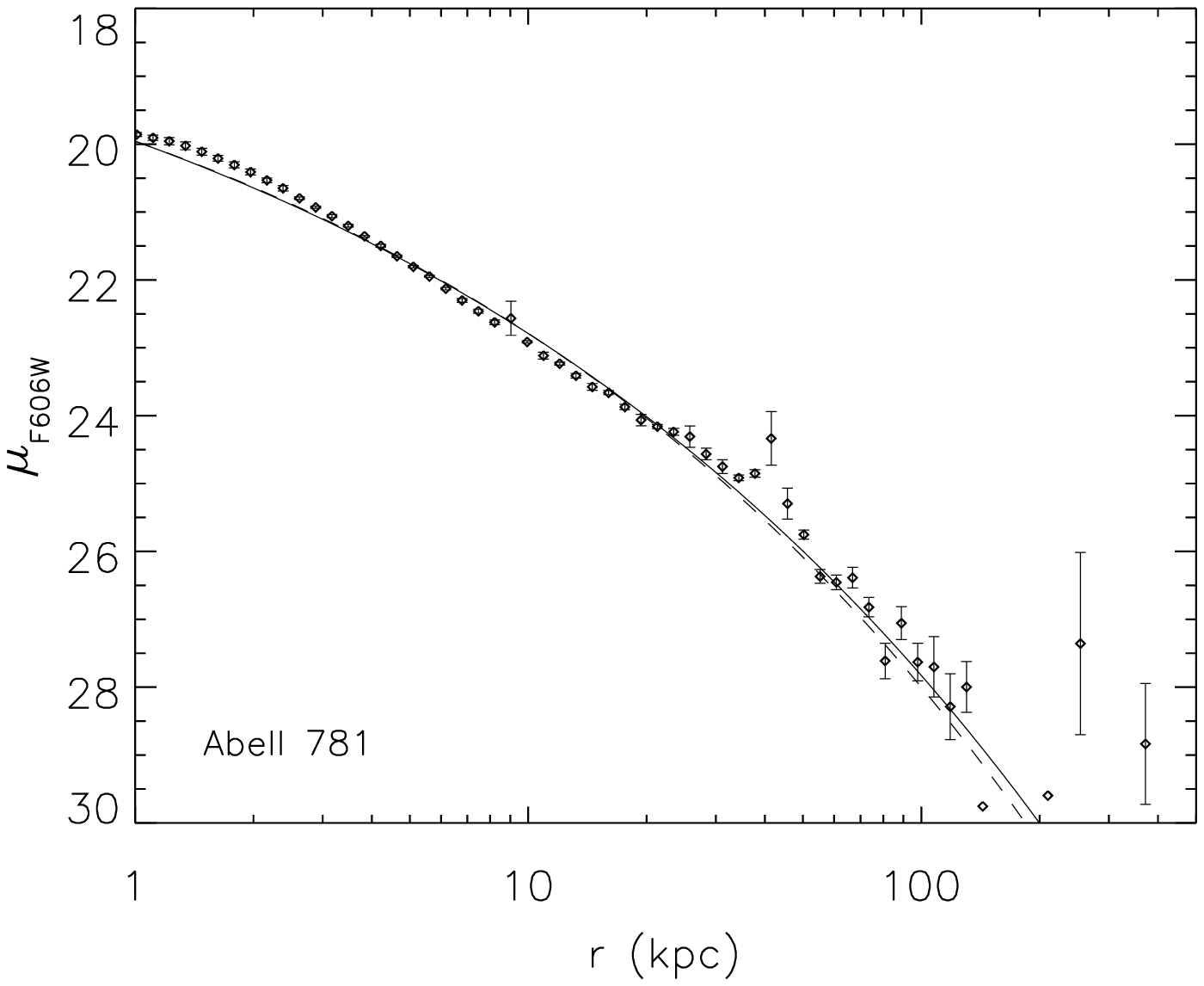} 
\includegraphics[scale=0.5]{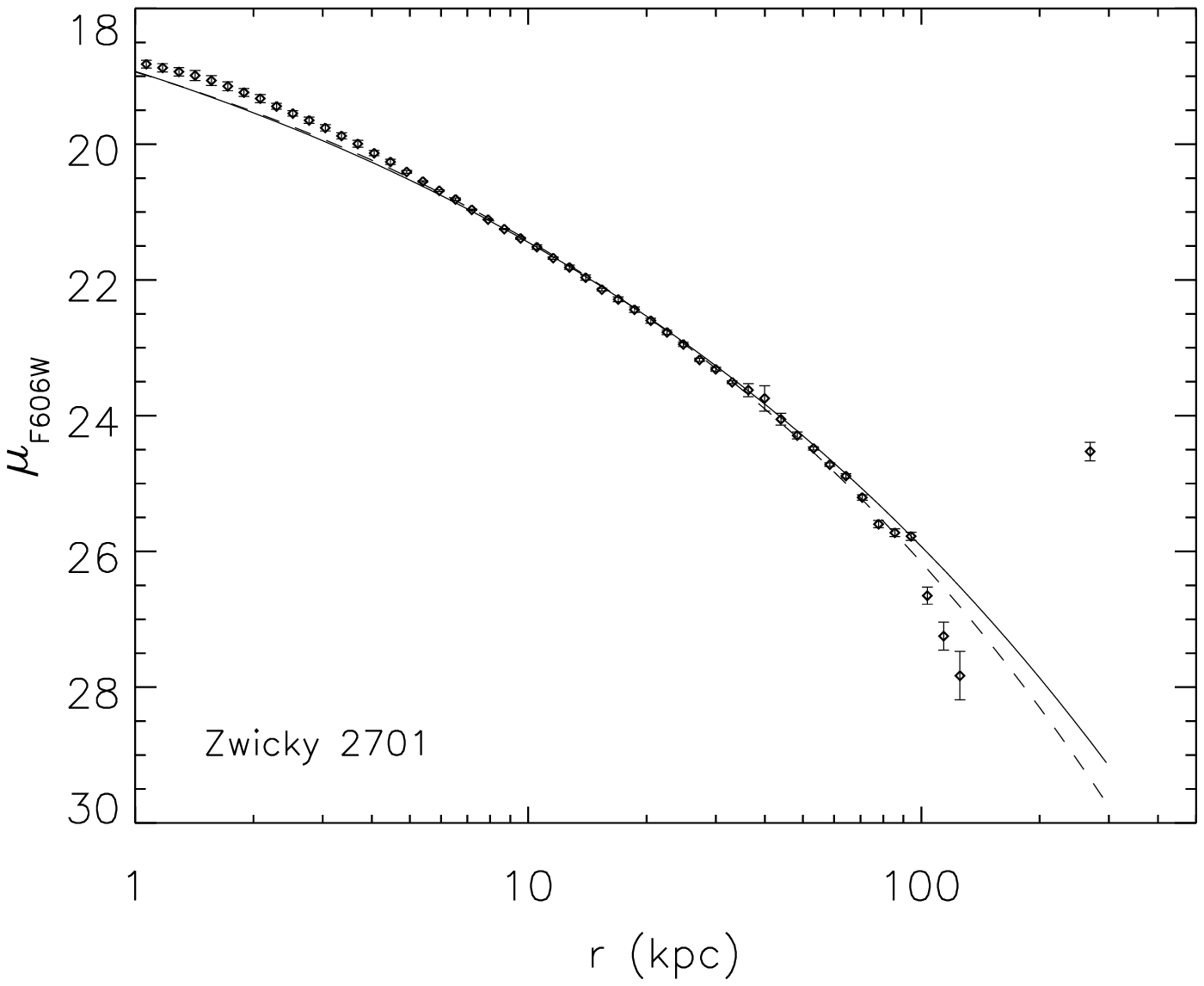} 
\includegraphics[scale=0.5]{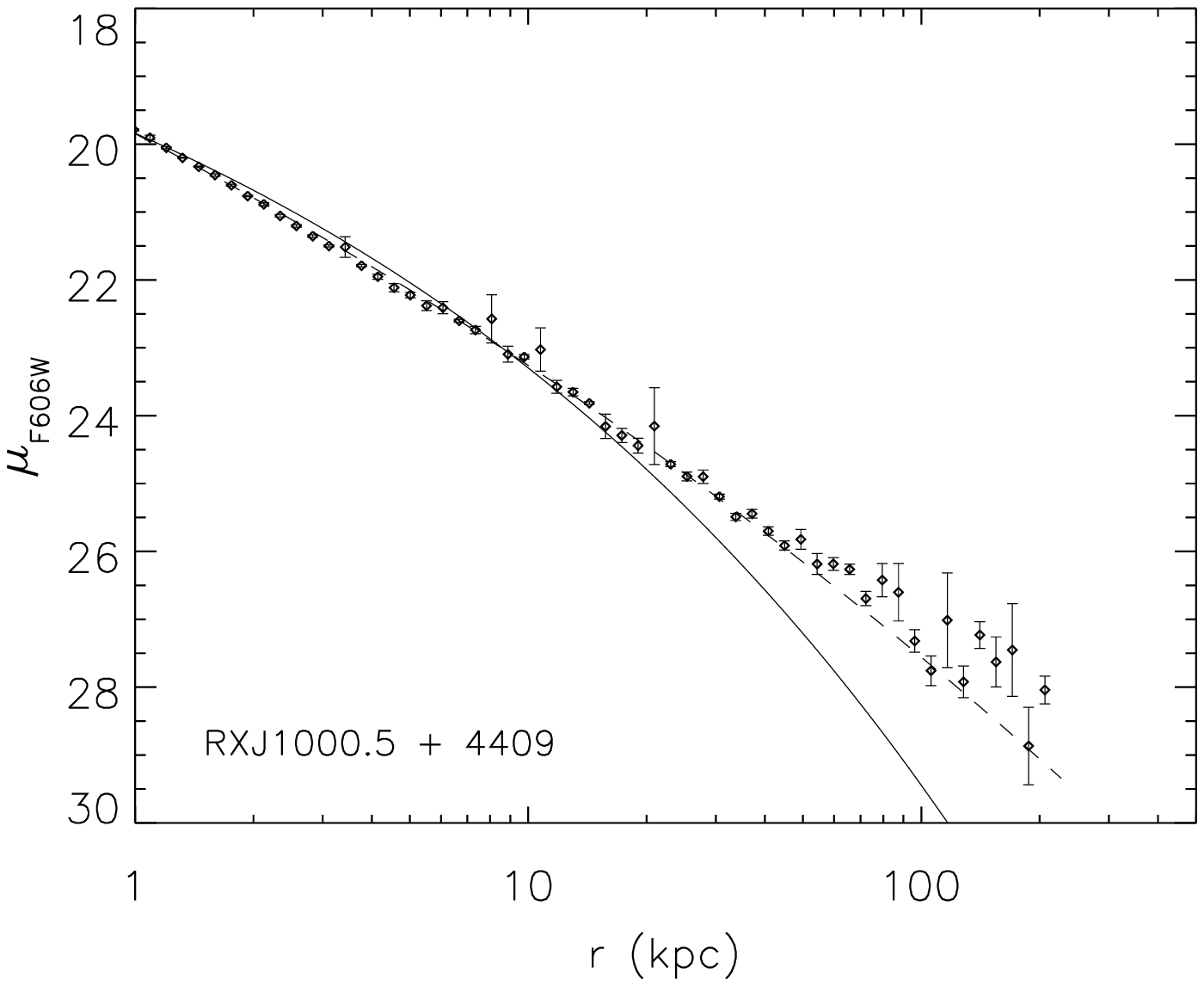} 
\includegraphics[scale=0.5]{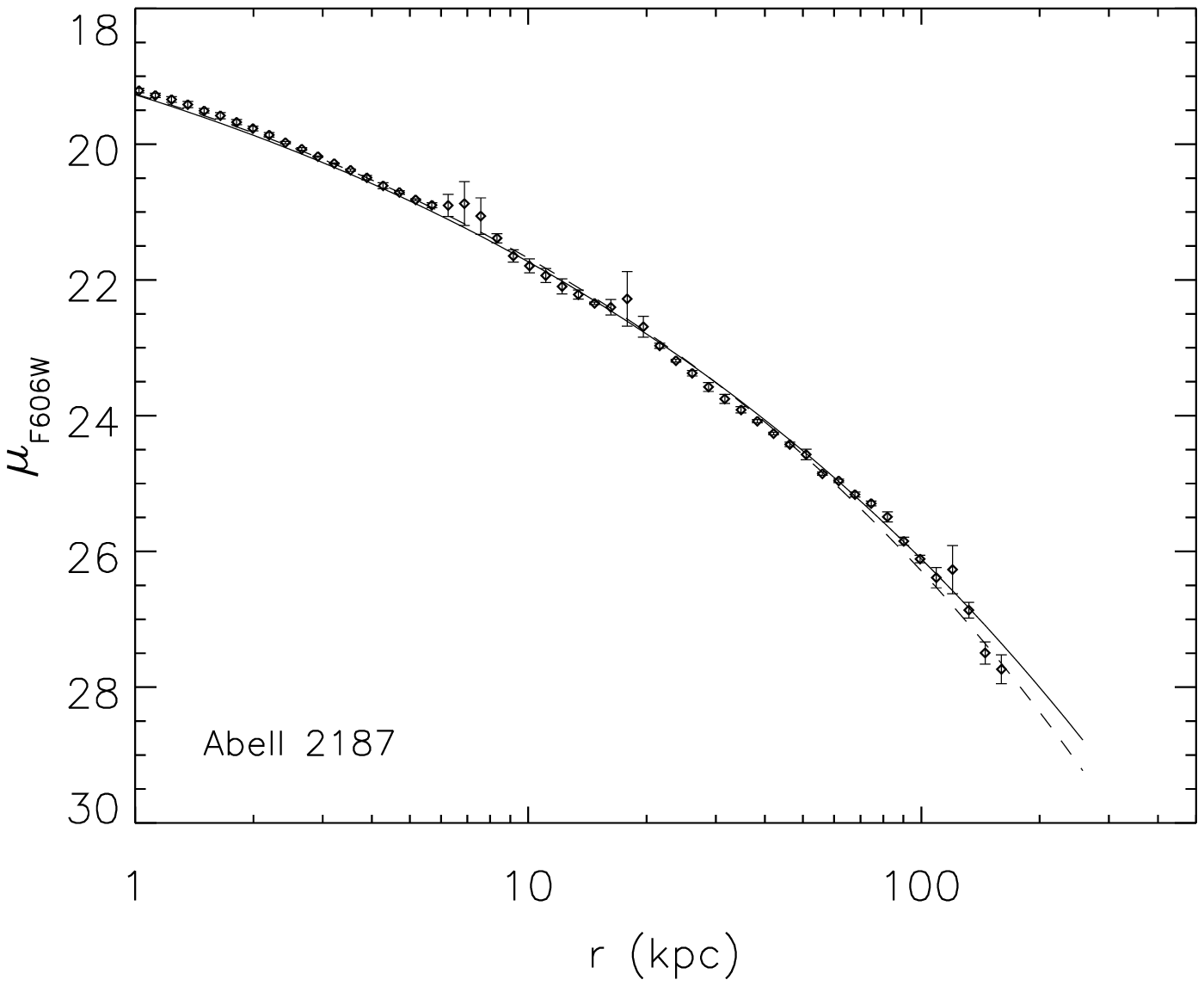} 
\caption{The surface brightness profiles for the individual LoCuSS BCGs 3.  de Vaucouleur and S\'{e}rsic profile fits plotted (solid and dashed lines respectively).}
   \label{fig:locuss3}
\end{figure*}

\begin{figure*}
   \centering
\includegraphics[scale=0.5]{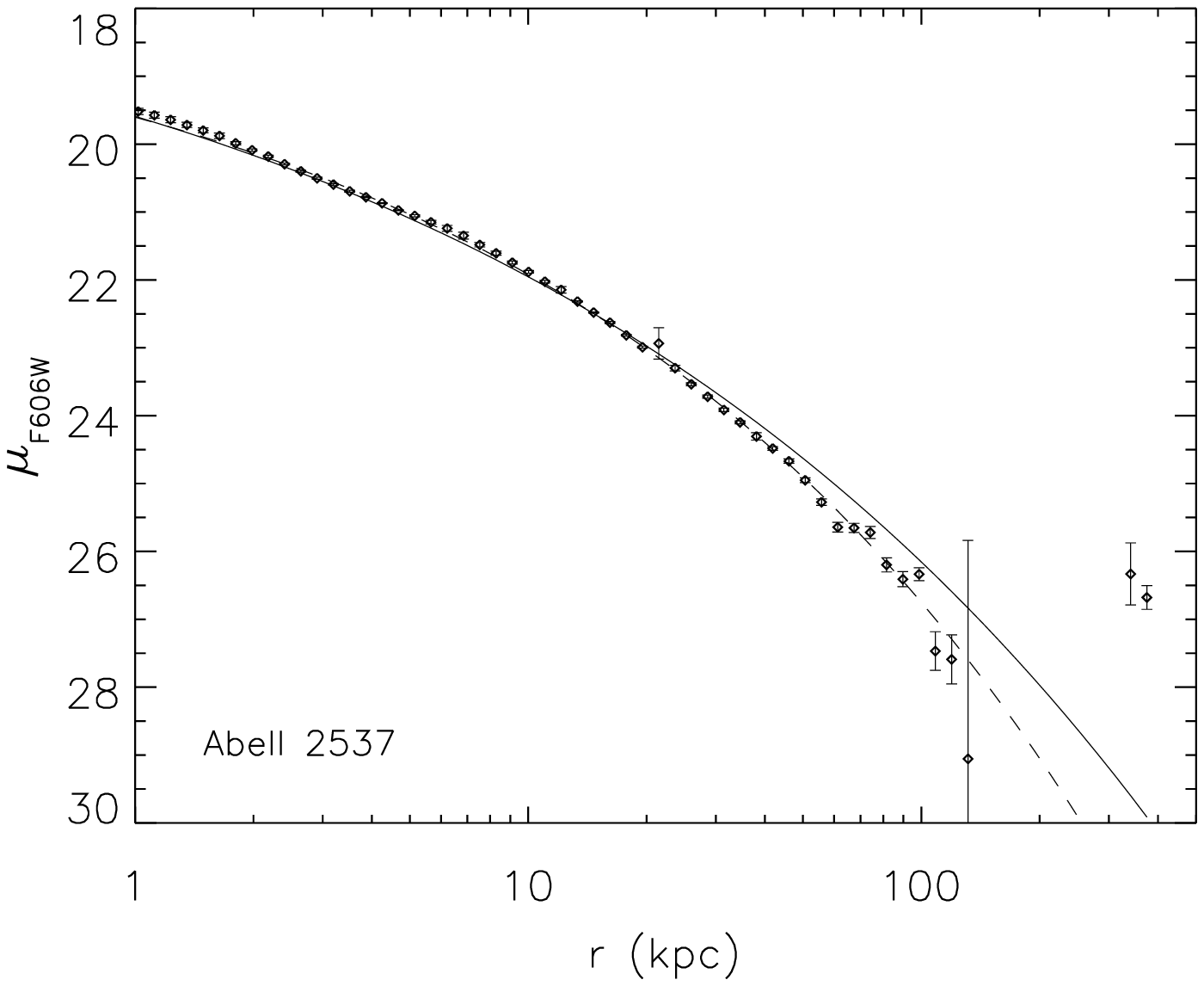} 

\caption{The surface brightness profiles for the individual LoCuSS BCGs 4.  de Vaucouleur and S\'{e}rsic profile fits plotted (solid and dashed lines respectively).}
   \label{fig:locuss4}
\end{figure*}

\begin{figure*}
   \centering
\includegraphics[scale=0.5]{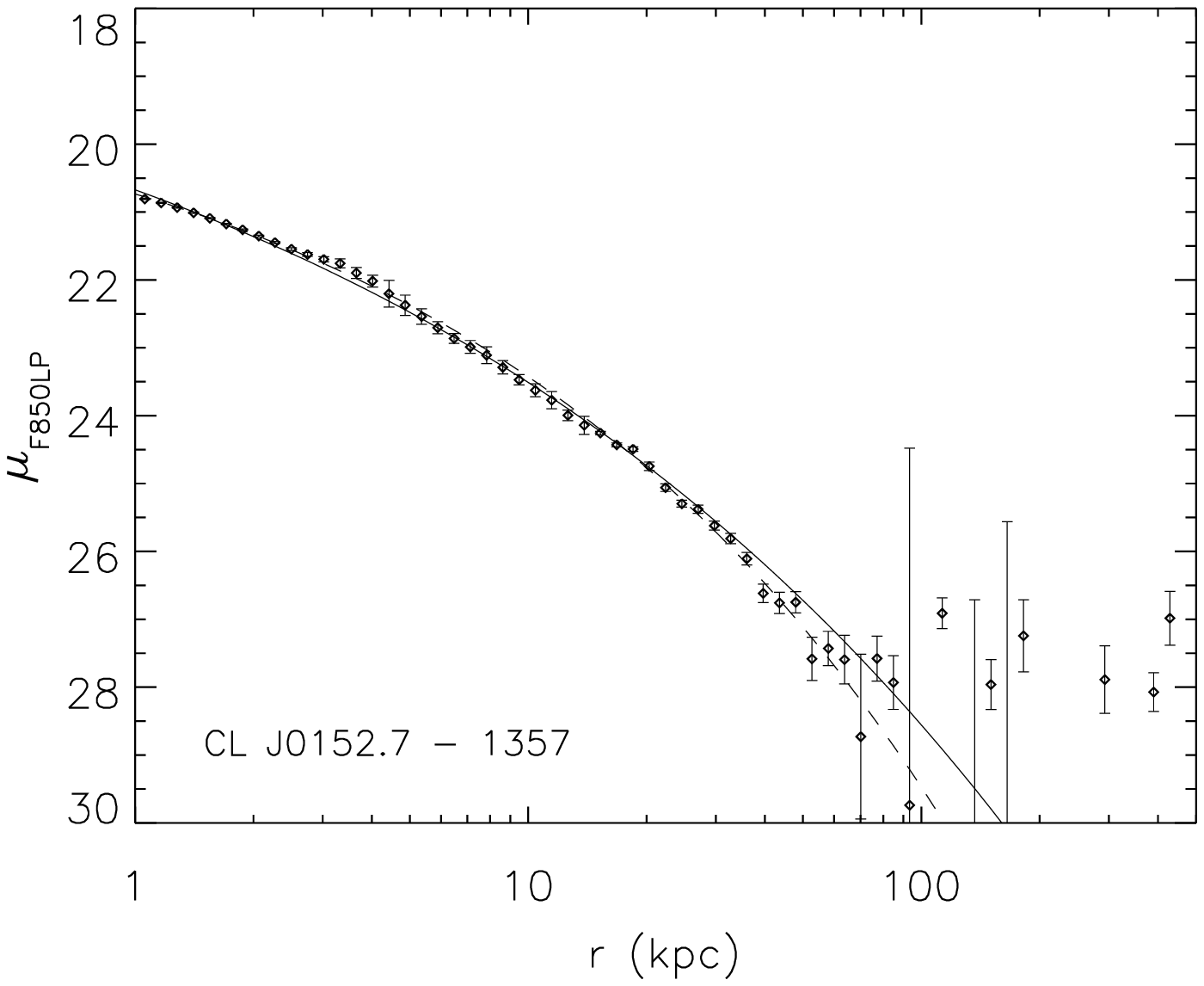} 
\includegraphics[scale=0.5]{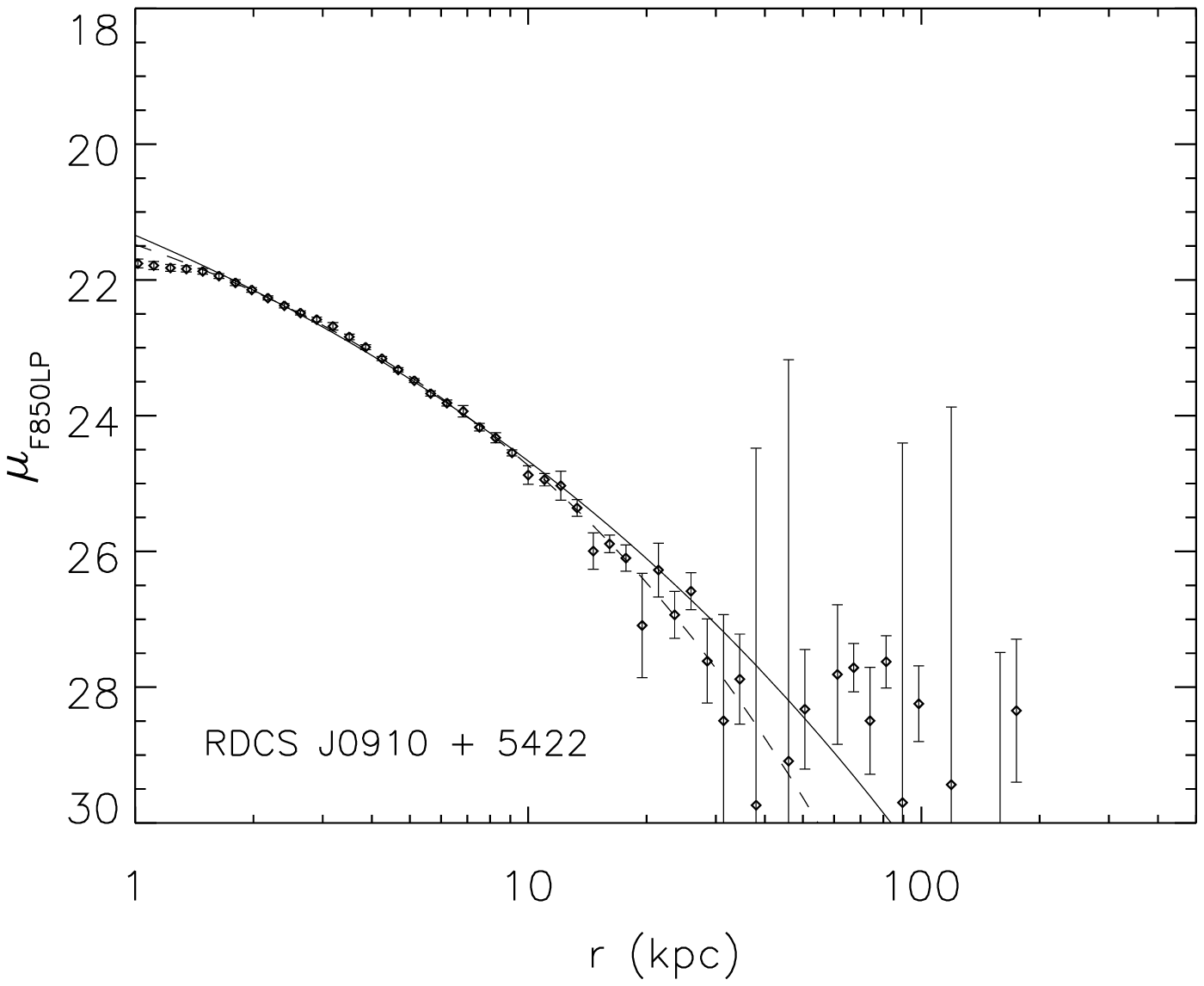} 
\includegraphics[scale=0.5]{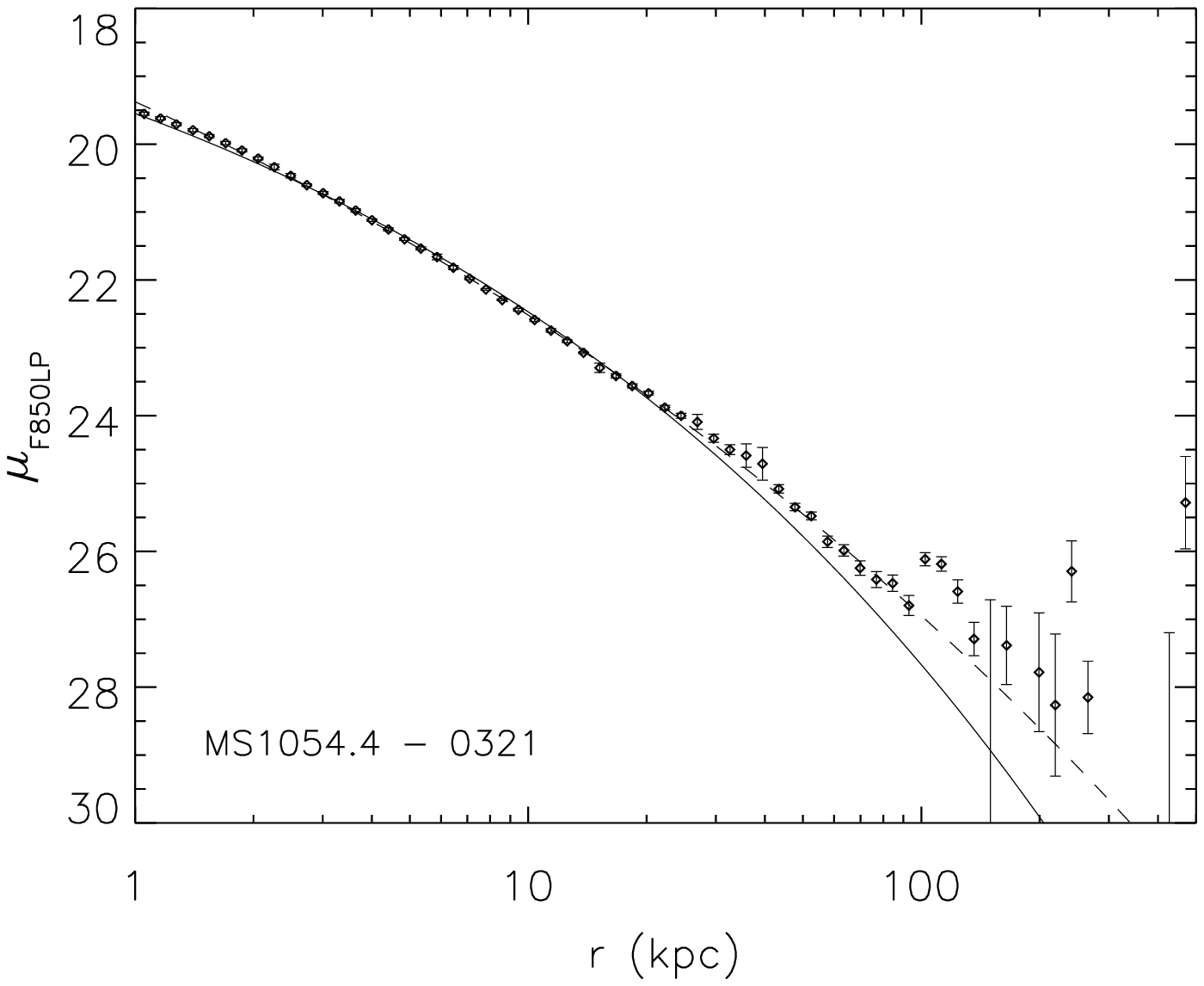} 
\includegraphics[scale=0.5]{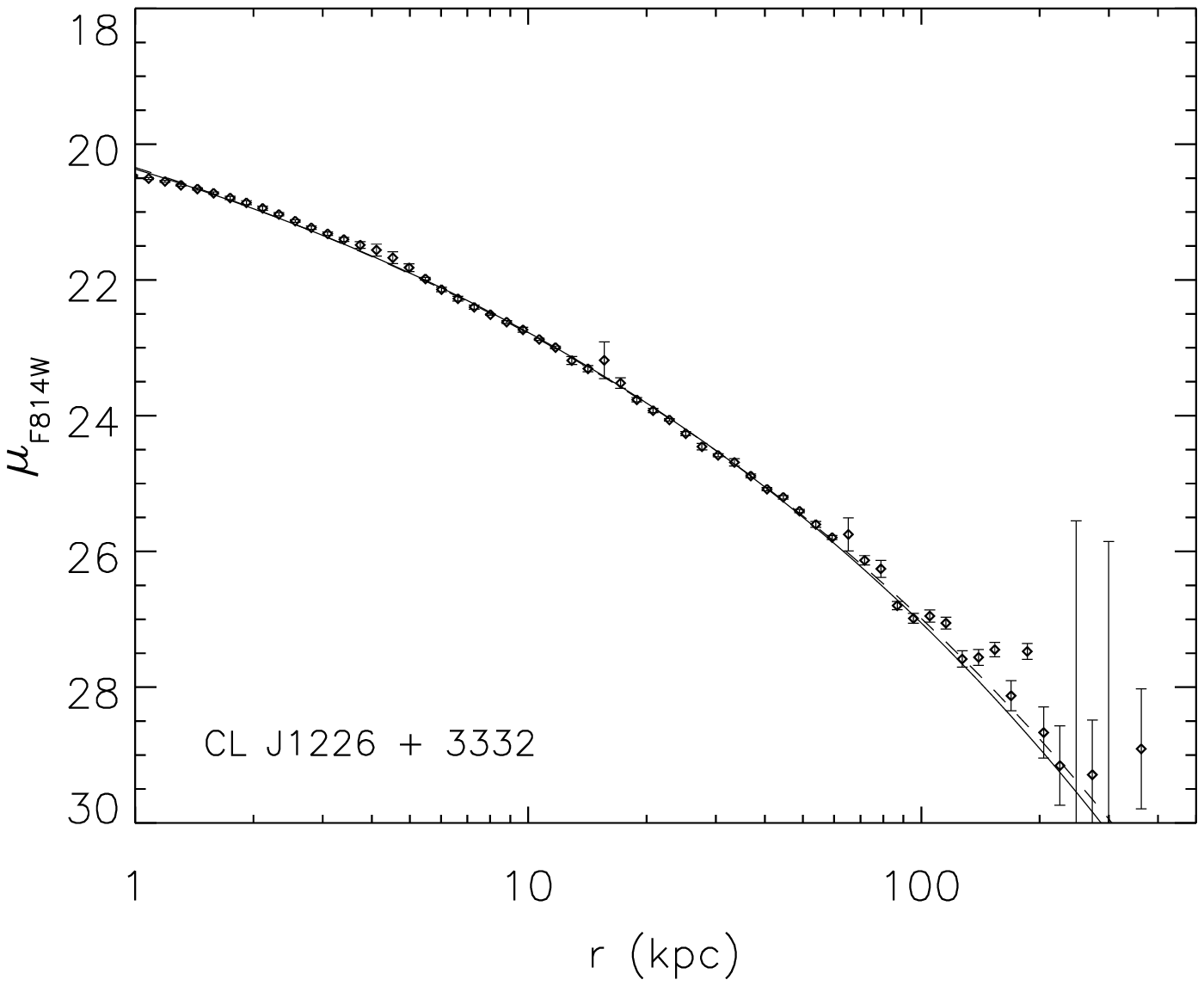} 
\includegraphics[scale=0.5]{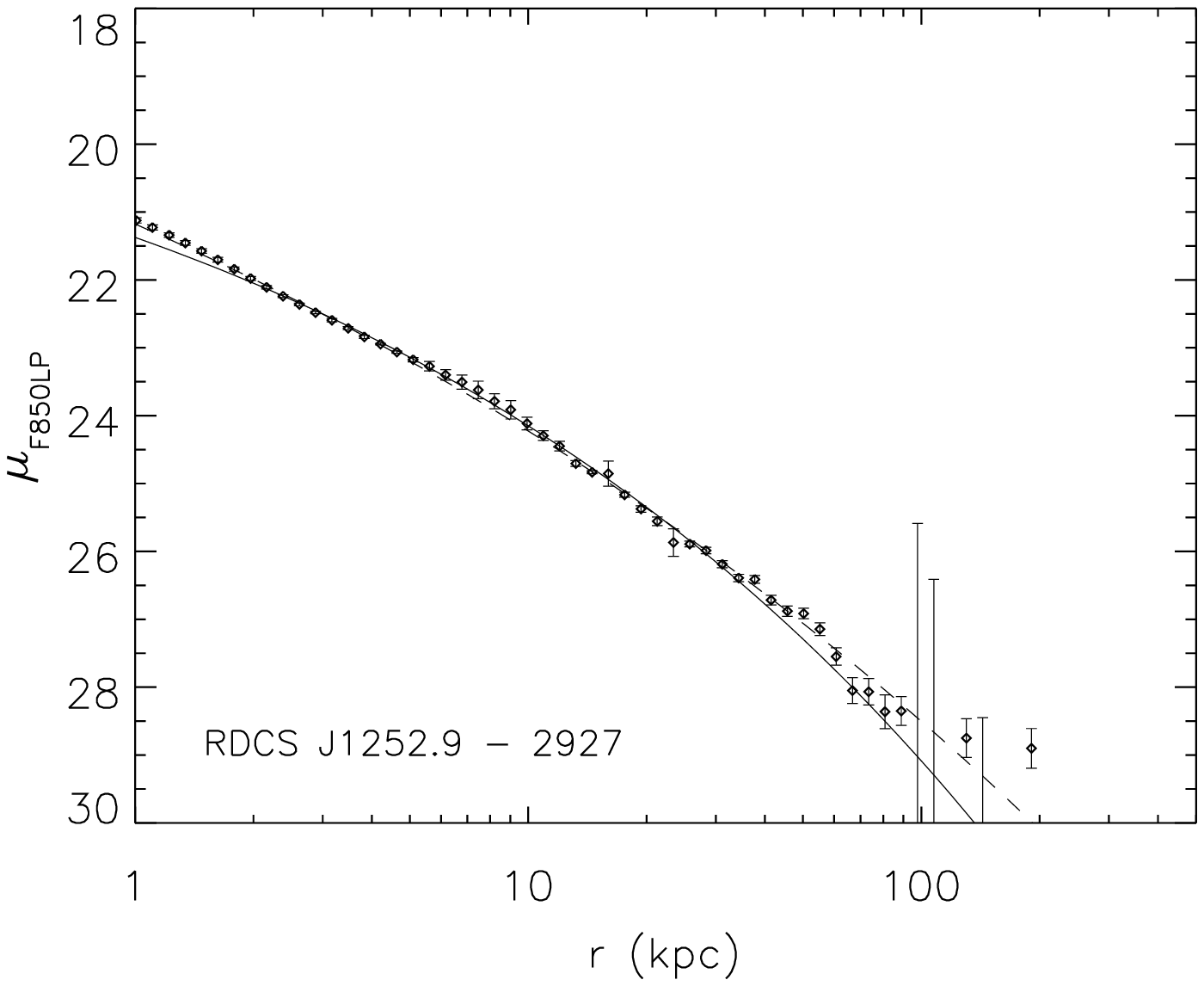} 

\caption{The surface brightness profiles for the individual high redshift BCGs with the de Vaucouleur and S\'{e}rsic profile fits plotted (solid and dashed lines respectively).}
   \label{fig:hizind}
\end{figure*}

\label{lastpage}

\end{document}